\documentclass[prl,twocolumn,nopacs,preprintnumbers,notitlepage,amsmath,amssymb,superscriptaddress]{revtex4-2}
\usepackage{amssymb}
\usepackage{amsmath}

\usepackage{color}
\usepackage[pdftex]{graphicx}
\usepackage{svg}

\usepackage{mathtools}

\usepackage{makecell}

\definecolor{darkblue}{rgb}{0,0,0.6}
\definecolor{darkred}{rgb}{0.6,0,0}
\usepackage[colorlinks=true,urlcolor=darkblue,citecolor=darkblue,linkcolor=darkred]{hyperref}

\usepackage{lipsum}

\usepackage{soul}

\usepackage{enumerate}
\usepackage{mathtools}
\usepackage{stmaryrd}

\usepackage{enumitem}

\usepackage{textcomp}
\usepackage{gensymb}

\newcommand{\moy}[1]{\left\langle #1 \right\rangle}

\newcommand{\XX}[0]{\boldsymbol{X}}

\newcommand{\ex}[1]{\mathrm{e}^{#1}}

\newcommand{\gt}[0]{\widetilde{g}}
\newcommand{\dd}[0]{\mathrm{d}}
\newcommand{\ii}[0]{\mathrm{i}}

\newcommand{\zz}[0]{\mathbf{0}}

\newcommand{\FF}[0]{\boldsymbol{F}}
\newcommand{\ee}[0]{\boldsymbol{e}}
\newcommand{\rr}[0]{\boldsymbol{r}}

\newcommand{\RR}[0]{\boldsymbol{R}}

\newcommand{\qq}[0]{\boldsymbol{q}}

\newcommand{\sigmaa}[0]{}

\begin{document}

\title{Microscopic theory for the diffusion of an active particle in a crowded environment}

\author{Pierre Rizkallah}
\affiliation{Sorbonne Universit\'e, CNRS, Laboratoire de Physico-Chimie des \'Electrolytes et Nanosyst\`emes Interfaciaux (PHENIX), 4 Place Jussieu, 75005 Paris, France}

\author{Alessandro Sarracino}
\affiliation{Dipartimento di Ingegneria, Universit\`a della Campania ”Luigi Vanvitelli”, 81031 Aversa (CE), Italy}

\author{Olivier Bénichou}
\affiliation{Sorbonne Universit\'e, CNRS, Laboratoire de Physique Th\'eorique de la Mati\`ere Condens\'ee (LPTMC), 4 Place Jussieu, 75005 Paris, France}

\author{Pierre Illien}
\affiliation{Sorbonne Universit\'e, CNRS, Laboratoire de Physico-Chimie des \'Electrolytes et Nanosyst\`emes Interfaciaux (PHENIX), 4 Place Jussieu, 75005 Paris, France}

\date{\today}

\begin{abstract}

We calculate the diffusion coefficient of an active tracer in a schematic crowded environment, represented as a lattice gas of passive particles with hardcore interactions. Starting from the master equation of the problem, we put forward a closure approximation that goes beyond trivial mean-field and provides the diffusion coefficient for an arbitrary density of crowders in the system. We show that our approximation is accurate for a very wide range of parameters, and that it correctly captures numerous nonequilibrium effects, which are the signature of the activity in the system.  In addition to the determination of the diffusion coefficient of the tracer, our approach allows us to characterize the perturbation of the environment induced by the displacement of the active tracer. Finally, we consider the asymptotic regimes of low and high densities, in which the expression of the diffusion coefficient of the tracer becomes explicit, and which we argue to be exact.

\end{abstract}

\maketitle

\emph{Introduction.---}
Many theoretical models of active particles have been introduced and studied during the past decades. They were proven to be particularly powerful to describe the dynamics of a large number of real systems, ranging from biological objects (molecular motors, bacteria, micro-swimmers, algae...) to artificial self-propelled particles such as active colloids \cite{Bechinger2016,Zottl2016a}. Among these models, run-and-tumble particles and active Brownian particles have attracted a lot of interest: in both cases, the particles self-propel with a fixed velocity, whose orientation changes randomly either abruptly or continuously, respectively. The dynamics of isolated or non-interacting active particles has been the subject of numerous recent studies \cite{Romanczuk2012,Tailleur2009,Cates2013,Malakar2018,Schnitzer1993, Martens2012, Kurzthaler2017, Basu2018, Basu2019}.

\begin{figure}[b]
    \centering
    \includegraphics[width=0.7\columnwidth]{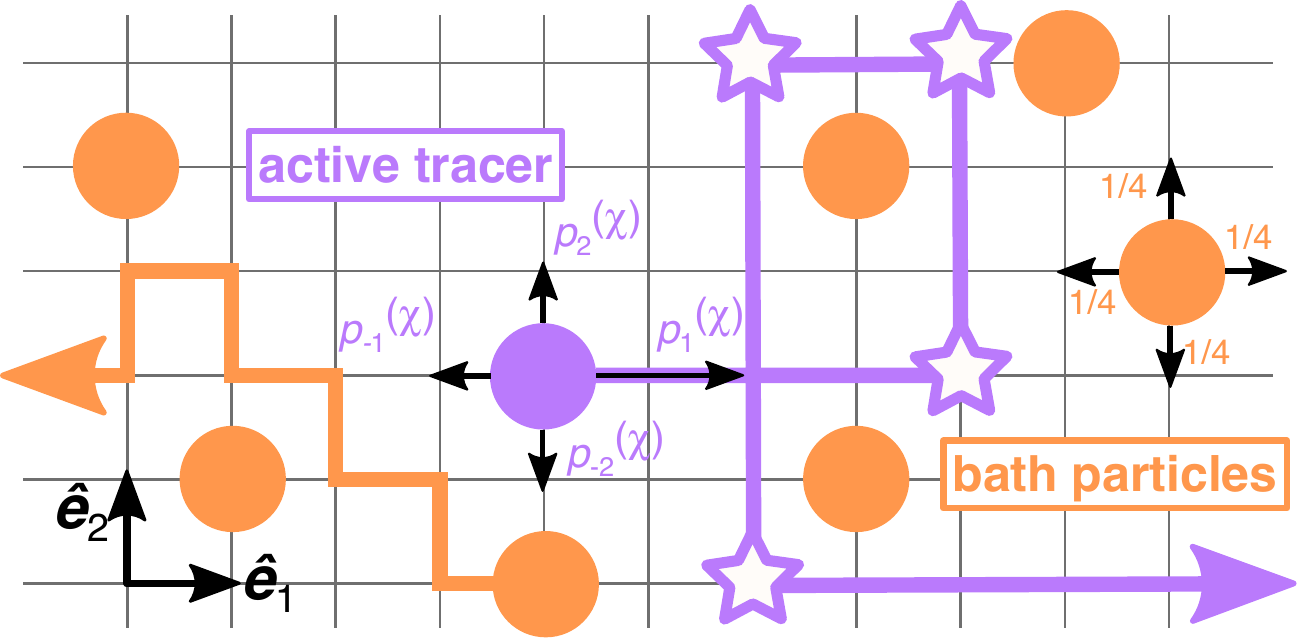}
    \caption{An active tracer performs a persistent random walk {(here, the active force initially points in direction $+\ee_1$, which corresponds to $\chi=1$)} in a bath of particles performing symmetric random walks (see text for notations). {The orientation of the bias of the tracer changes randomly, and the re-orientation events are represented by stars}.}
    \label{fig:model}
\end{figure}

Beyond single-particle properties, the dynamics of active particles when they interact with each other has attracted a lot of attention, and was shown to display numerous surprising effects, such as large-scale collective motion \cite{Vicsek2012}, clustering, or phase separation in the absence of attractive interactions \cite{Bechinger2016,Cates2015}. In addition, it is crucial to understand the interactions between active particles and complex environments. Indeed, the transport of many biological objects takes place under crowded conditions, such as motor proteins inside a cell \cite{Conway2012a} or bacteria in porous materials \cite{Licata2016}.
So far, the transport of active particles in frozen disordered environments was studied through experiments (on living \cite{Bhattacharjee2019,  Makarchuk2019,{Sipos2015}, Brun-Cosme-Bruny2019, Guidobaldi2014} and synthetic \cite{Morin2017} microswimmers) and theoretical approaches (essentially numerical) \cite{Chepizhko2013, Tailleur2009,Kaiser2012, Reichhardt2014,Bijnens2021,Chepizhko2020, Volpe2017,Zeitz2016,Jakuszeit2019,{MariniBettoloMarconi2017},{Caprini2020,Caprini2018}}.


The case of dynamic disorder, which has received much less attention, is however particularly relevant, since thermal fluctuations generally affect the environment as well as the tracer \cite{Hofling2013}. Models involving tracers in environments of mobile obstacles (Fig. \ref{fig:model}) have therefore been employed to describe situations of biological interest \cite{Saxton1987, Schmit2009, Dorsaz2010}. For the case of a \emph{passive} tracer, the celebrated theory by Nakazato and Kitahara  \cite{Nakazato1980} (see also \cite{Tahir-Kheli1983a,vanBeijeren1985a}) gives an expression of the corresponding diffusion coefficient as a function of the density of crowders, in a continuous-time description. Due to the many-body nature of the problem, this expression is approximate but has been shown to be exact in the low and high density regimes, and offers very good quantitative estimates for arbitrary density, as soon as the environment is mobile enough \cite{Tahir-Kheli1983a,vanBeijeren1985a}. The case of an \emph{active} tracer in a dynamic environment 
has been the subject of only a few theoretical studies of particles evolving on a lattice {(see however \cite{Reichert2020} for a very recent mode-coupling approach in continuous space)}, which focused mainly on the low-density limit of the problem, with a discrete-time description, with a tracer that never jumps sideways from the direction of propulsion, and with a specific dynamics \cite{Bertrand2018a}. Particular interactions between particles (third-neighbor exclusion) have also been studied through numerical simulations and mean-field approximations \cite{Chatterjee2019}. A generic analytical framework, that would allow the calculation of the diffusivity of an active tracer in a dynamic environment for a wide range of parameters, and in particular for arbitrary density, is  missing. {Indeed, although discrete space models for the diffusion of tracers have attracted a lot of attention and have proven particularly efficient to characterise dynamics in crowded environments, there is no continuous-time lattice model that incorporates both the effect of activity and that of crowding at arbitrary density, and that quantifies tracer-bath correlations.}

In this Letter, we provide a microscopic theory for the diffusion coefficient of an active tracer in  a crowded environment on a lattice, at arbitrary density and activity. Adopting a standard continuous-time dynamics and starting from the master equation describing the joint probability distribution for the position of the tracer and the configuration of its environment, we resort to a closure approximation and calculate the diffusion coefficient of the active tracer in terms of the bath density profiles, and of tracer-bath correlation functions. Importantly, in addition to the determination of the diffusion coefficient of the tracer, our approach allows us to calculate the perturbation of the environment due to the displacement of the active tracer, and the space dependence of the correlations between the tracer position and the bath occupation numbers. Finally, the expression for the diffusion coefficient becomes explicit in the low- and high-density regimes, in which we claim that our closure approximation becomes exact.

\emph{Model.---} We consider an active tracer in a crowded and dynamic environment (Fig. \ref{fig:model}). The bath particles (of density $\rho$), and the tracer evolve on a $d$-dimensional cubic lattice, whose spacing is taken equal to $1$. As opposed to discrete-time descriptions \cite{Bertrand2018a}, the system evolves here in continuous time, which is the natural and usual way to describe systems with site-blocking effects, both in one dimension as in (Asymmetric) Simple Exclusion Processes \cite{Chou2011, Mallick2015} and in higher dimensions \cite{Nakazato1980, Tahir-Kheli1983a, vanBeijeren1985a}. Note that the dynamics of a biased tracer is known to be significantly affected by the choice of dynamics (discrete-time or continuous-time) \cite{Benichou2013g}. The bath particles perform symmetric nearest-neighbor random walks (with characteristic time $\tau^*$), and the tracer  performs a random walk (with characteristic time $\tau$) biased in the direction of an active force whose orientation changes randomly.  The variable $\chi \in \{ \pm 1,\dots,\pm d \}$ is the `state' of the tracer, i.e. the direction {in which the active force points}. The tracer switches  from a state $\chi$ to any other state $\chi' \neq \chi$ with rate $\frac{\alpha}{2d\tau^*}$, where $\alpha$ is dimensionless. The persistence time is then $\tau_\alpha = \frac{2d\tau^*}{\alpha}$.  We denote by $p_\mu^{(\chi)}$ the probability for the tracer to jump in direction $\mu \in \{ \pm 1,\dots,\pm d \}$ when it is in state $\chi$. Given that the active force is in a random direction $\chi$, we choose $p_\mu^{(\chi)} \propto\exp[F_A \ee_\chi \cdot\ee_\mu/2]$ with an appropriate normalization (where $\ee_{\pm1}, \dots, \ee_{\pm d} $ are the lattice unit vectors and we use the notation $\ee_{-\mu} = -\ee_\mu$). The active force $F_A$ is easily related to the velocity of the tracer in the absence of crowding interactions   $  v_0 = (p^{(1)}_1-p^{(1)}_{-1})/{\tau}  $ \footnote{$v_0$ is the analogous of the propulsion velocity in usual continuous-space models of active Brownian or run-and-tumble particles. Here, given that the particle evolves on a lattice, $v_0$ is bounded by $1/\tau$. Note that, given our choice of $p_\mu^{(\chi)}$, the active force $\FF$ controls $v_0$ but also controls the magnitude of the fluctuations in the direction perpendicular to $\FF$.}. 
The dynamics of the tracer is a lattice representation of run-and-tumble dynamics, which is a central model in the theory of active matter, and which has been widely used to describe the transport and diffusion of bacteria, see for instance \cite{Schnitzer1993}. Finally, all the particles evolve on the lattice with the restriction that there can only be one particle per site, which mimics hardcore interactions.

The state of the system at time $t$ is described by $P_\chi (\RR,\eta;t)$, which is the joint probability to find the tracer in state $\chi$, at site $\RR$, with the lattice in configuration $\eta=\{\eta_{\rr}\}$, where $\eta_{\rr}=1$ if site $\rr$ is occupied by a bath particle and $0$ otherwise. The master equation obeyed by the joint tracer-bath probability is :
\begin{equation}
2d\tau^*\partial_t P_\chi(\RR,\eta;t) = \mathcal{L}_\chi P_\chi - \alpha P_\chi +\frac{\alpha}{2d-1} \sum_{\chi'\neq\chi} P_{\chi'}
\label{mastereq}
\end{equation}
where $\mathcal{L}_\chi$ is the evolution operator in state $\chi$ and is given in the Supplemental Material (SM) \cite{SM}. {It accounts for the diffusion of the tracer and of the bath particles, whereas the last two terms of Eq. \eqref{mastereq} account for the random changes in the orientation of the active force.}

At $t=0$, we assume that all the directions of the active force are equally likely, in such a way that the mean position of the tracer particle remains zero, and that at any time all states have the same probability $\frac{1}{2d}$. We are interested in the fluctuations of the tracer position along one direction, for instance $X_t=\XX_t\cdot\ee_1$ {(where $\XX_t=X_t \ee_1+\sum_{k=2}^d X_t^{(k)} \ee_k$)}. Multiplying the master equation by $X_t^2$ and averaging yields an expression for the time derivative of $\moy{X_t^2}$, where $\moy{\cdot}$ denotes the average over the position of the tracer, its state, and the configuration of the lattice. The long-time diffusion coefficient of the tracer, defined as $D\equiv \lim_{t\to\infty}\frac{1}{2} \frac{\dd}{\dd t} \moy{X_t^2}$, can be written under the form \cite{SM}
\begin{align}
&D  =  \dfrac{1}{4d\tau}\sum_{\chi} \sum_{\epsilon=\pm1} \left\{ p_\epsilon^{(\chi)} \left[1-k_{\epsilon}^{(\chi)}\right] -2\epsilon p_\epsilon^{(\chi)} \gt_{\epsilon}^{(\chi)} \right\}  \nonumber \\
&  + \frac{2d-1}{2d} \frac{\tau^*}{\tau^2 \alpha}  \sum_{\chi}  \left\lbrace \sum_{\epsilon=\pm1} \epsilon  p_\epsilon^{(\chi)} \left[1-k_{\epsilon}^{(\chi)}\right] \right\rbrace^2. \label{eq:diffcoeff_def}
\end{align}
This expression involves the density profiles in the frame of reference of the tracer $k_{\rr}^{(\chi)} = \moy{\eta_{\XX_t+\rr}}_\chi$ and tracer-bath cross-correlations functions $\gt_{\rr}^{(\chi)} = \langle \eta_{\XX_t+\rr} (X_t - \moy{X_t}_\chi)\rangle_\chi$, where $\moy{\cdot}_\chi = 2d \sum_{\RR,\eta} \cdot P_\chi(\RR,\eta;t)$ denotes the average conditioned on state  $\chi$ \footnote{Note that the calculation of $D$ requires the calculation of $X^\infty_\chi$ -- the average position of the tracer conditioned on state $\chi$ in the stationary state -- whose expression is given in SM \cite{SM}}.

\emph{Decoupling approximation.---}
The equations governing $k_{\rr}^{(\chi)}$ and $\gt_{\rr}^{(\chi)}$, which are obtained by multiplying the master equation [Eq. \eqref{mastereq}] respectively by $\eta_{\XX_t+\rr}$ and  $X_t \eta_{\XX_t+\rr}$,   are not closed and involve higher-order correlation functions, whose evolution equations involve even higher-order correlation functions, and so on. The resulting infinite hierarchy of equations is closed by the following mean-field-type approximation:
$\moy{\eta_{\rr}\eta_{\rr'}}_{\chi} \simeq k_{\rr}^{(\chi)}k_{\rr'}^{(\chi)}$ and $\moy{\delta X_t \eta_{\rr}\eta_{\rr'}}_{\chi} \simeq k_{\rr}^{(\chi)} \gt_{\rr'}^{(\chi)} +  k_{\rr'}^{(\chi)} \gt_{\rr}^{(\chi)}$,
which is obtained by writing each random variable as $x=\moy{x}+\delta x$ and neglecting terms of order 2 and 3 in the fluctuations. Note that this goes beyond trivial mean-field, in which the mean occupation of the lattice sites would be assumed to be uniform and equal to $\rho$. This approximation has been successfully applied to study the velocity \cite{Benichou2014} and diffusivity \cite{Illien2017c} of a driven tracer (limit of $\alpha\to0$) and has been shown to become exact in the low- and high-density regimes \cite{Benichou2018}.

We obtain the following equations for $h_{\rr}^{(\chi)} \equiv k_{\rr}^{(\chi)} - \rho$ (defined in such a way that $\lim_{|\rr|\to\infty} h_{\rr}=0$) and $\gt_{\rr}^{(\chi)}$ (we adopt the convention $h_{\zz}^{(\chi)} =\gt_{\zz}^{(\chi)} = 0$):
\begin{align}
&2d\tau^*\partial_t h_{\rr}^{(\chi)} =  (\tilde{L}^{(\chi)} + \textstyle\sum_\nu 
A_\nu^{(\chi)}\delta_{\rr,\ee_\nu})   h_{\rr}^{(\chi)} \nonumber \\
&+\textstyle\sum_\nu\delta_{\rr,\ee_\nu} \rho(A_\nu-A_{-\nu}) 
- \alpha h_{\rr}^{(\chi)} + \frac{\alpha}{2d-1}\sum_{\chi' \neq \chi} h_{\rr}^{(\chi')}, \label{eq:h_Euler}\\
&2d\tau^*\partial_t \gt_{\rr}^{(\chi)}  =  (\tilde{L}^{(\chi)}+ \textstyle\sum_\nu  
A_\nu^{(\chi)}\delta_{\rr,\ee_\nu}) \gt_{\rr}^{(\chi)} + \mathcal{G}^{(\chi)}h_{\rr}^{(\chi)}- \alpha \gt_{\rr}^{(\chi)} \nonumber\\
&+ \frac{\alpha}{2d-1}\textstyle\sum_{\chi' \neq \chi} \gt_{\rr}^{(\chi')} +\textstyle\sum_\nu \delta_{\rr,\ee_\nu}\left[ \left(A_{-\nu}^{(\chi)} - 1 \right) \rho(\ee_{\nu} \cdot \ee_1) \right.  \nonumber\\
&\left. - \dfrac{2d\tau^*}{\tau}\left( p_\nu^{(\chi)} \gt_{\ee_{\nu}}^{(\chi)}  \left( h_{\ee_{\nu}}^{(\chi)} + \rho\right) - \rho p_{-\nu}^{(\chi)} \gt_{\ee_{-\nu}}^{(\chi)}\right) \right], \label{eq:g_Euler}
\end{align}
where we define $A_\mu^{(\chi)} \equiv 1 + \frac{2d\tau^*}{\tau} p_\mu^{(\chi)} [1-k_{\ee_\mu}^{(\chi)}] $, the operator $\tilde{L}^{(\chi)}$ acting on a test function $f_{\rr}$ as $\tilde{L}^{(\chi)}f_{\rr} \equiv  \sum_\mu A_\mu^{(\chi)} \left(f_{\rr + \ee_{\mu}} - f_{\rr} \right)$. The operator $\mathcal{G}^{(\chi)}$ is defined in SM \cite{SM}.
The sums over $\mu$ and $\nu$ implicitly run over all $2d$ directions of the lattice. 
Eqs. \eqref{eq:h_Euler} and \eqref{eq:g_Euler} constitute one of the main results of our Letter: within our closure approximation, these equations allows the determination of the quantities $h_{\rr}^{(\chi)}$ and $\gt_{\rr}^{(\chi)}$, and therefore of the diffusion coefficient of the tracer through Eq. \eqref{eq:diffcoeff_def}, for an arbitrary set of parameters, and in particular for an arbitrary density of crowders $\rho$.

\emph{Resolution.---}
The  resolution of Eqs. \eqref{eq:h_Euler} and \eqref{eq:g_Euler} relies on the translational invariance of the system, enabling us to use Fourier transforms to invert the discrete-space differential operator. We define the following Fourier transforms, where the sum on $\rr$ runs over lattice sites: $H^{(\chi)}(\qq;t) \equiv \sum_{\rr} \ex{\ii\qq\cdot\rr} h_{\rr}^{(\chi)}(t) $ and $
G^{(\chi)}(\qq;t) \equiv \sum_{\rr} \ex{\ii\qq\cdot\rr} \gt_{\rr}^{(\chi)}(t) $. The Fourier transforms of Eqs. \eqref{eq:h_Euler} and \eqref{eq:g_Euler} are given in the SM \cite{SM}.
In the stationary state, these equations are written under the form $  \boldsymbol{M}(\qq) \boldsymbol{H}(\qq) + \boldsymbol{R_H}(\qq) =  0$ and $ \boldsymbol{M}(\qq) \boldsymbol{G}(\qq) + \boldsymbol{R_G}(\qq) =  0$, where we define the $2d$-dimensional vectors $\boldsymbol{H}(\qq) \equiv \left( H^{(1)}(\qq), H^{(-1)}(\qq),\dots \right)$ and $\boldsymbol{G}(\qq) \equiv\left( G^{(1)}(\qq),G^{(-1)}(\qq),\dots  \right)$. $\boldsymbol{R_H}$ depends only on $h_{\ee_\mu}^{(\chi)}$, and $\boldsymbol{R_G}$ depends on $h_{\ee_\mu}^{(\chi)}$ and $\gt_{\ee_\mu}^{(\chi)}$. $\boldsymbol{M}(\qq)$ is a matrix such that $[\boldsymbol{M}(\qq)]_{\chi\chi} = -\alpha + \sum_{\mu}   \left(\ex{-\ii\sigmaa q_\mu} -1\right)A_\mu^{(\chi)} $ and the off-diagonal terms are all $\alpha/(2d-1)$, where we use the shorthand notation $q_\mu \equiv \qq\cdot \ee_\mu$. The matrix $\boldsymbol{M}(\qq)$ is invertible (for $\qq \neq \zz$), and we deduce $\boldsymbol{H}(\qq) = -\boldsymbol{M}(\qq)^{-1}\boldsymbol{R_H}(\qq)$ and $\boldsymbol{G}(\qq) = -\boldsymbol{M}(\qq)^{-1}\boldsymbol{R_G}(\qq)$.
Then, by performing inverse Fourier transforms, we get a system satisfied by the quantities $h_\mu^{(\chi)}$ and $\gt_\mu^{(\chi)}$ \cite{SM}. This {system} makes it possible to calculate, within our approximation scheme, the diffusion coefficient for an arbitrary density of particles, with arbitrary values of the parameters $\tau$, $\tau^*$, $F_A$, and $\alpha$.

\begin{figure}
\begin{center}
\includegraphics[width = 0.45\columnwidth]{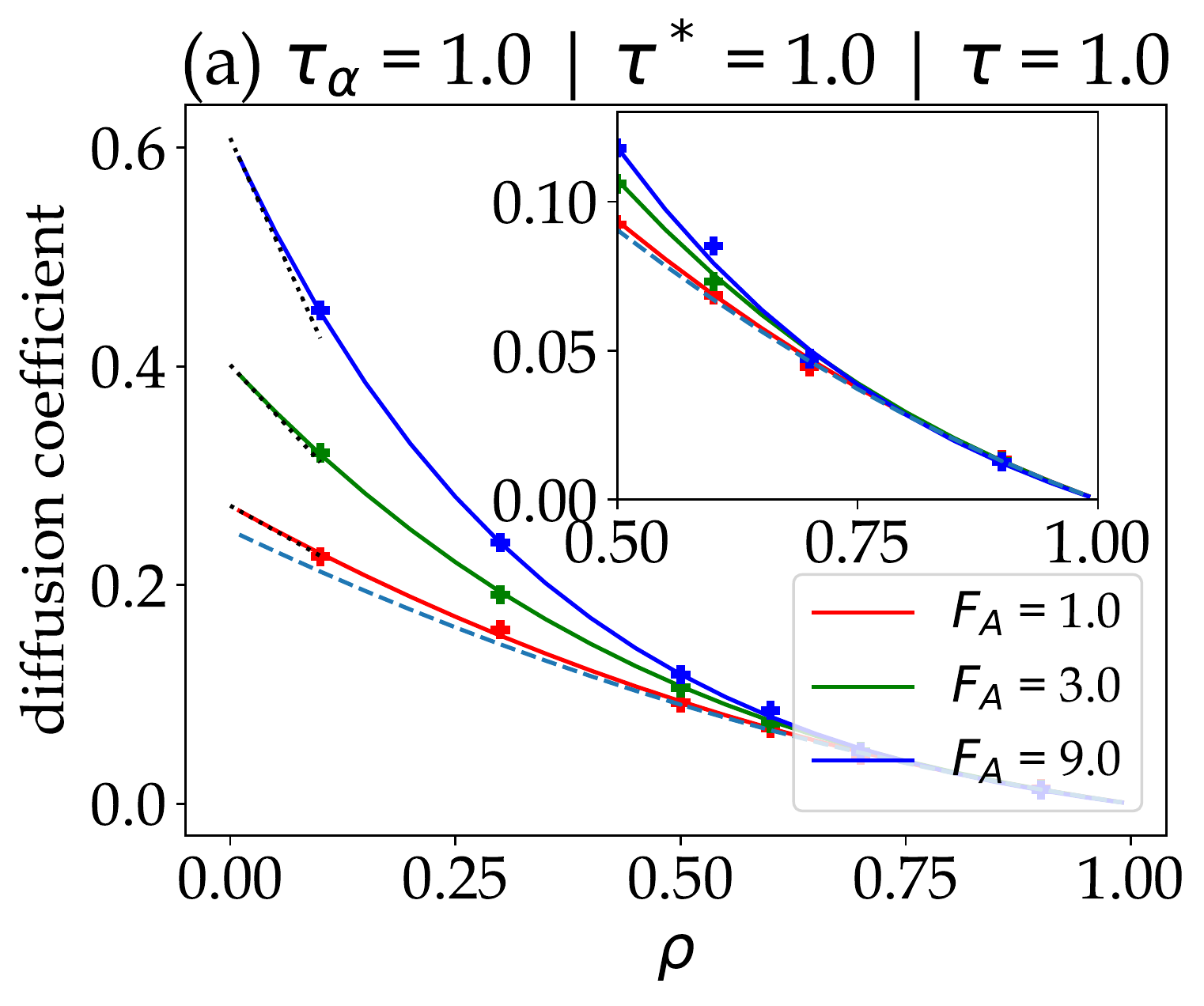}
\includegraphics[width = 0.45\columnwidth]{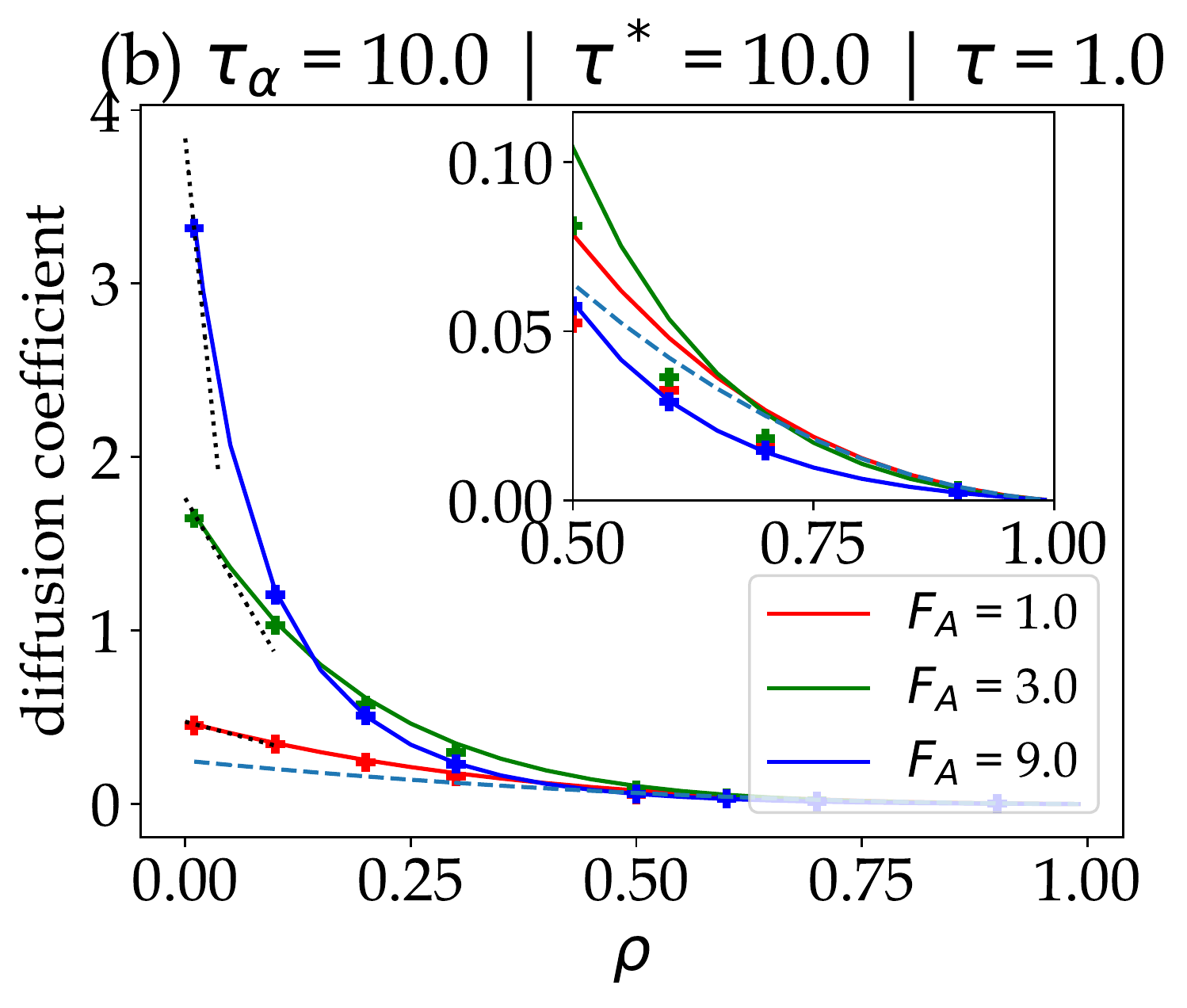}
\includegraphics[width = 0.45\columnwidth]{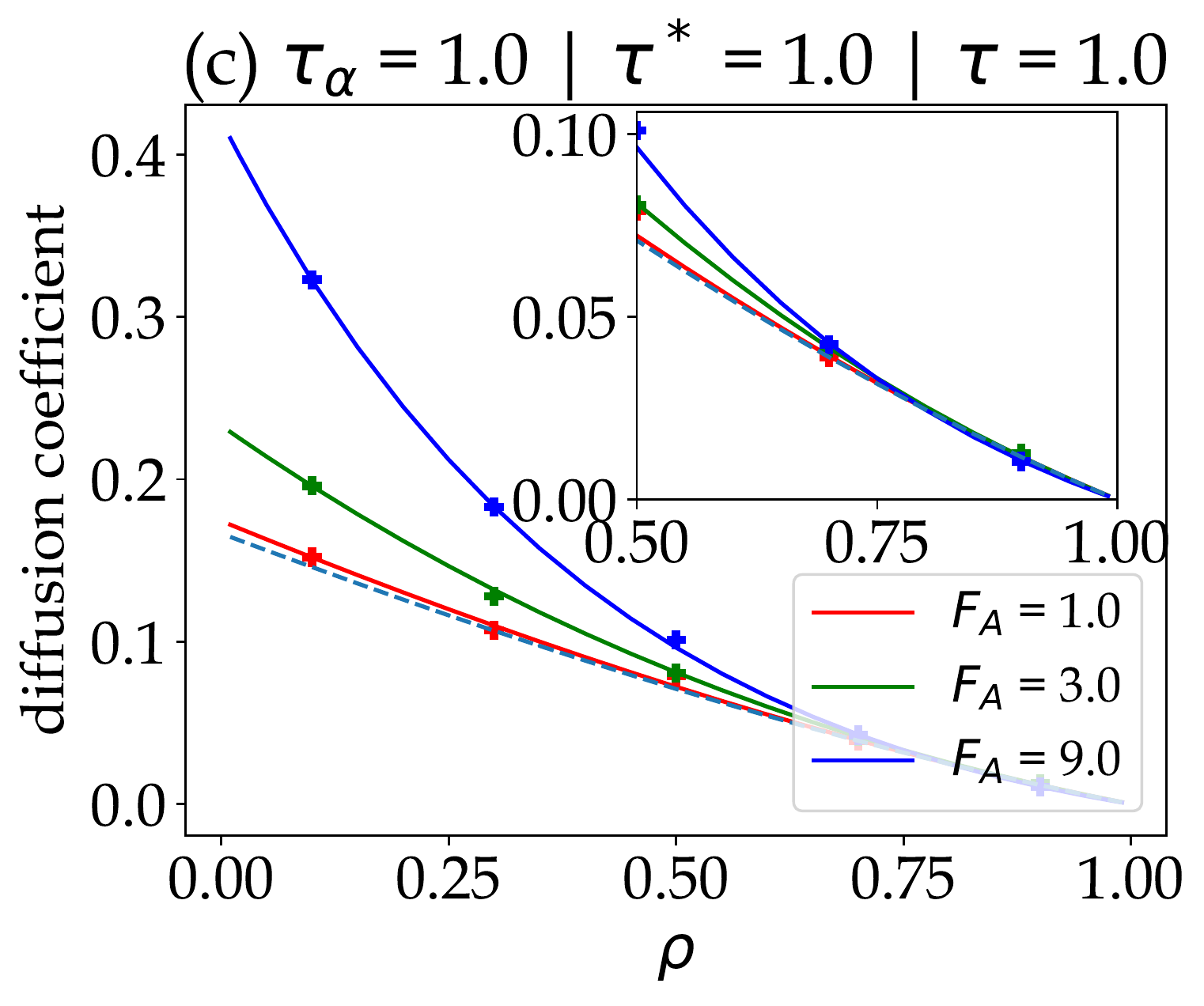}
\includegraphics[width = 0.45\columnwidth]{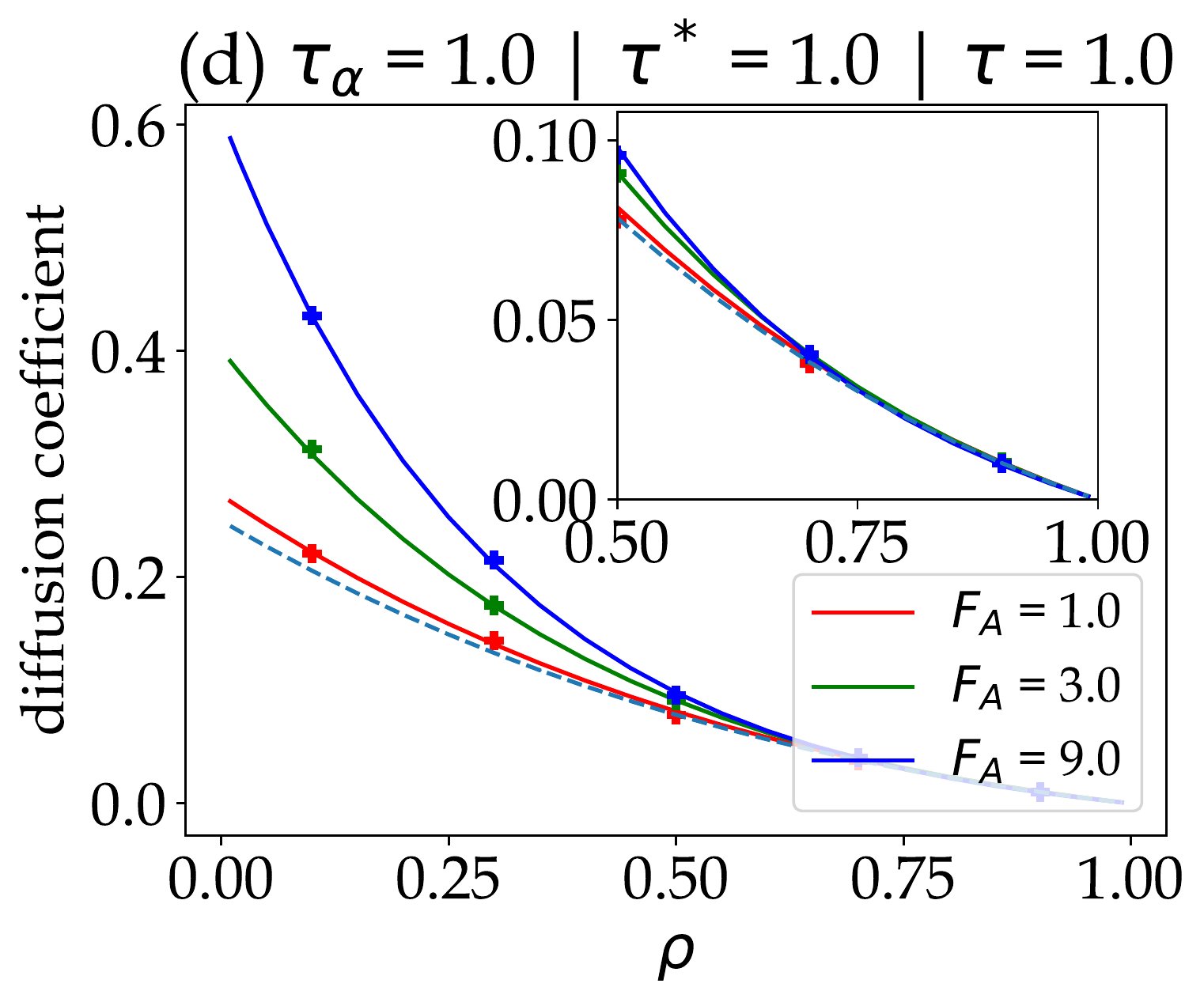}
\caption{Diffusion coefficient of an active tracer on a 2D lattice (a,b), a 3D lattice (c) and a 2D capillary of width $L=3$ (d), as a function of the density $\rho$, for several values of the active force $F_A$ and the persistence time $\tau_\alpha$. Symbols: Monte-Carlo simulations \cite{SM}. 
 Solid lines: analytical approach [Eqs. \eqref{eq:diffcoeff_def}, \eqref{eq:h_Euler}, and \eqref{eq:g_Euler}]. Dotted lines: asymptotic expansion in the low-density regime. Dashed lines: case of a passive tracer \cite{Nakazato1980}.  
} 
\label{fig:drho_2D}
\end{center}
\end{figure}

We first give the solution for a 2D infinite lattice. We compute {numerically} the values of $h_\mu^{(\chi)}$ and $\gt_\mu^{(\chi)}$ in the stationary state from Eqs. \eqref{eq:h_Euler} and \eqref{eq:g_Euler} \cite{SM}, and deduce the value of the diffusion coefficient using Eq. \eqref{eq:diffcoeff_def}. We study the dependence of $D$ on the density of particles on the lattice $\rho$, for different values of $\tau$, $\tau^*$, $\tau_\alpha$ and $F_A$. Fig. \ref{fig:drho_2D} displays very good agreement between Monte Carlo simulations and our decoupling approximation. {As in the theory for a passive tracer \cite{Nakazato1980}, the accuracy of our decoupling approximation improves when the crowding environment is more mobile (typically $\tau^*/\tau \lesssim 10$) or when the dimension of the lattice is higher.} In the case when there is no propulsion ($F_A=0$), our approximation matches the result by Nakazato and Kitahara \cite{Nakazato1980}, which provides an explicit expression of the diffusion coefficient as a function of the density, and which is recalled in SM \cite{SM}. Our result can therefore be seen as a generalization of this classical result on tracer diffusion in lattice gases to the case of an active particle. Note also that in the limit of $\alpha\to0$, we retrieve the results obtained previously for the velocity and the diffusion coefficient of a passive driven tracer \cite{Illien2017c}.

This calculation can easily be extended to other lattice geometries, provided that they remain translation-invariant. More specifically, we consider the case of a 2D stripe-like lattice (infinite in one direction and finite of width $L$ with periodic boundary conditions in the other direction), which schematically mimics narrow channels and confined systems, and of a 3D infinite lattice (Fig.~\ref{fig:drho_2D}).

\begin{figure}
\includegraphics[width=0.50\columnwidth]{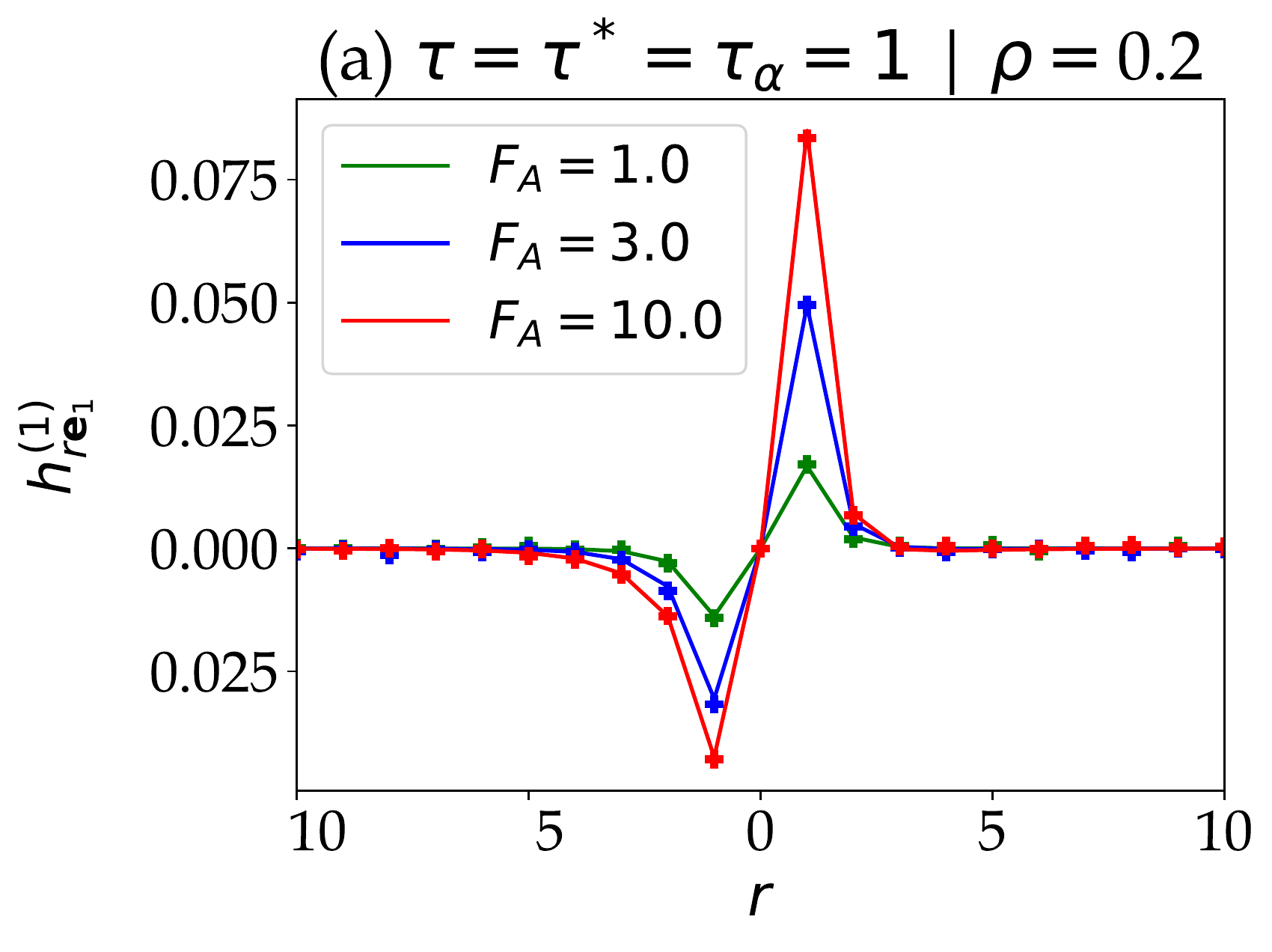}
\includegraphics[width=0.48\columnwidth]{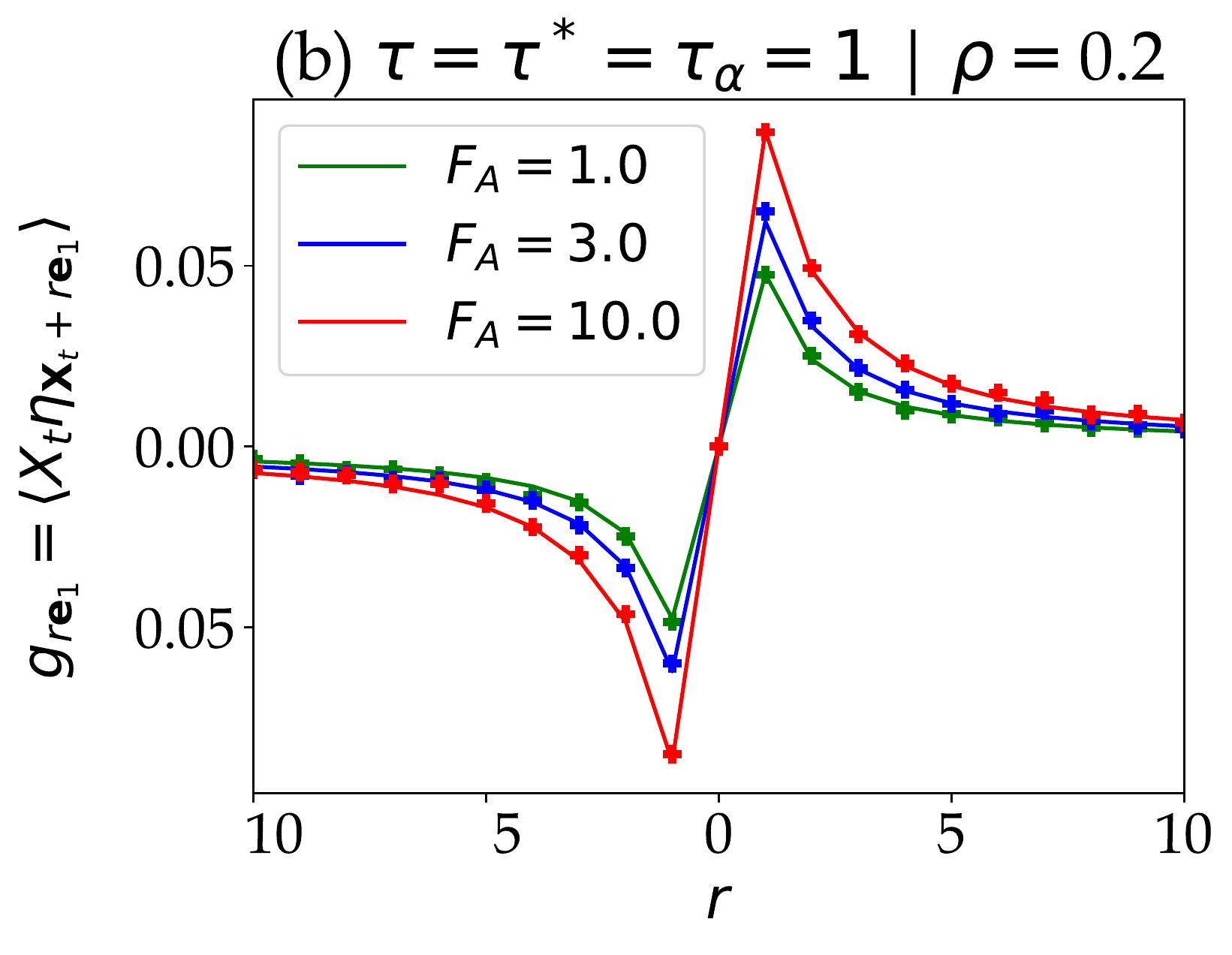}
\caption{(a) Density profiles (conditioned on the activity being in state $\chi =1$) and (b) tracer-bath correlation functions (averaged on all the states) as a function of the distance to the tracer, on a 2D lattice. {The values of the forces $F_A$ respectively correspond to transition probabilities $p_1^{(1)} = 0.39$, $0.67$ and $0.99$. }
{Symbols: Monte-Carlo simulations. Solid lines: analytical approach.}
}
\label{fig:profiles_h}
\end{figure}

 Finally, we emphasize that our approach allows us to go beyond the determination of the only diffusion coefficient of the tracer, and gives access to the perturbation induced by the activity of the tracer on its environment. More precisely, we calculate the complete space dependence of the density profiles $h_{\rr}^{(\chi)}$ and of the cross-correlation functions $\gt_{\rr}^{(\chi)}$ by performing inverse Fourier transforms of $H^{(\chi)}(\qq;t)$ and $G^{(\chi)}(\qq;t)$  (Fig. \ref{fig:profiles_h}). These quantities unveil the interplay between the displacement of the active tracer and the response of its environment -- an aspect out-of-reach of previous descriptions \cite{Bertrand2018}. In particular, we observe {{and quantify}} an accumulation of bath particles in front of the tracer and a depletion behind it. This local anisotropy of the environment of the tracer is a direct consequence of its activity, and is fully accounted for by our approach. {We provide an analytical framework to quantify the effect of active tracers on their environments, which is a key problem of active matter, with promising applications to use active tracers as microrheological probes \cite{Lozano2019}.}

\begin{figure}
\includegraphics[width=0.45\columnwidth]{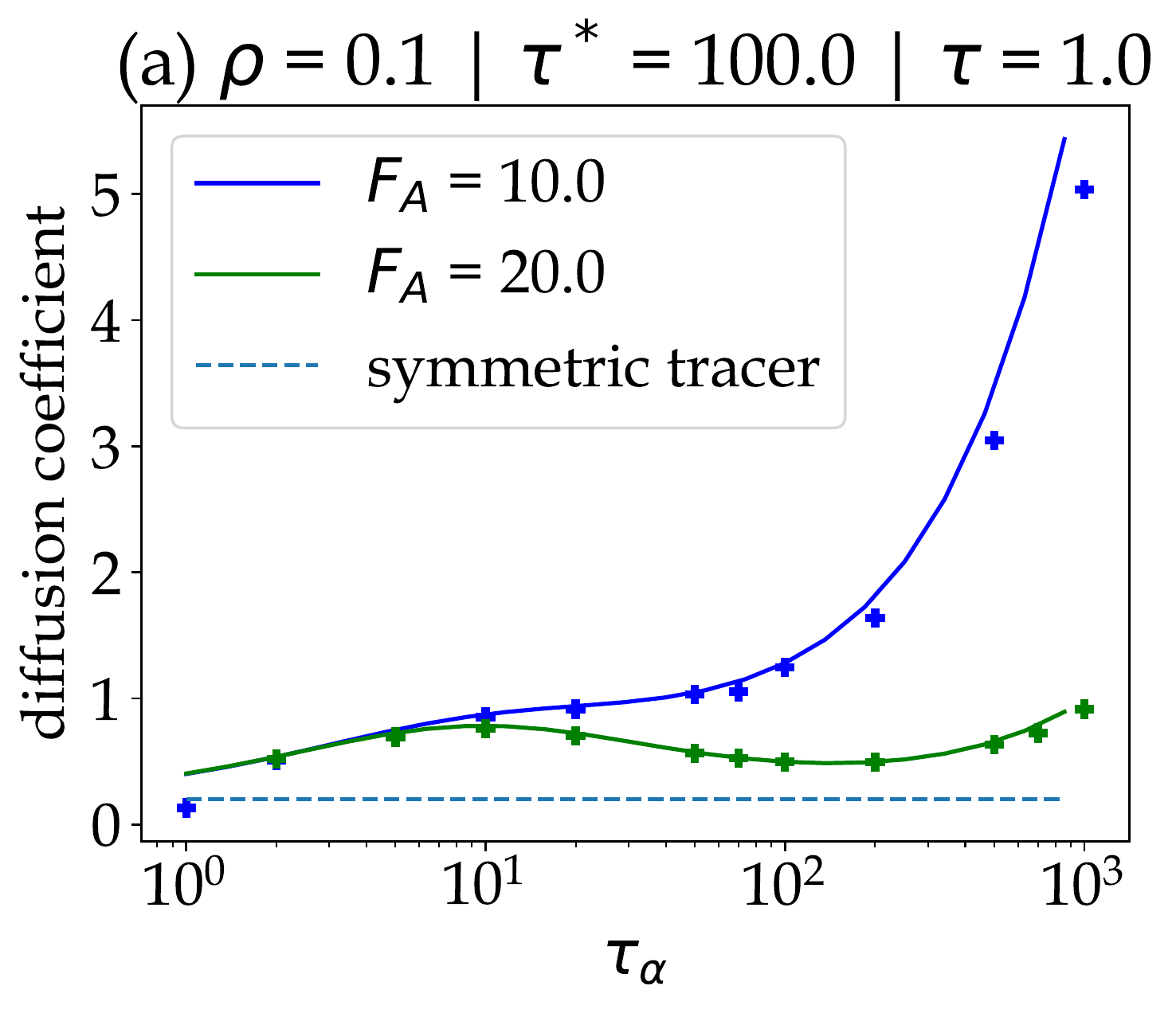}
\includegraphics[width=0.51\columnwidth]{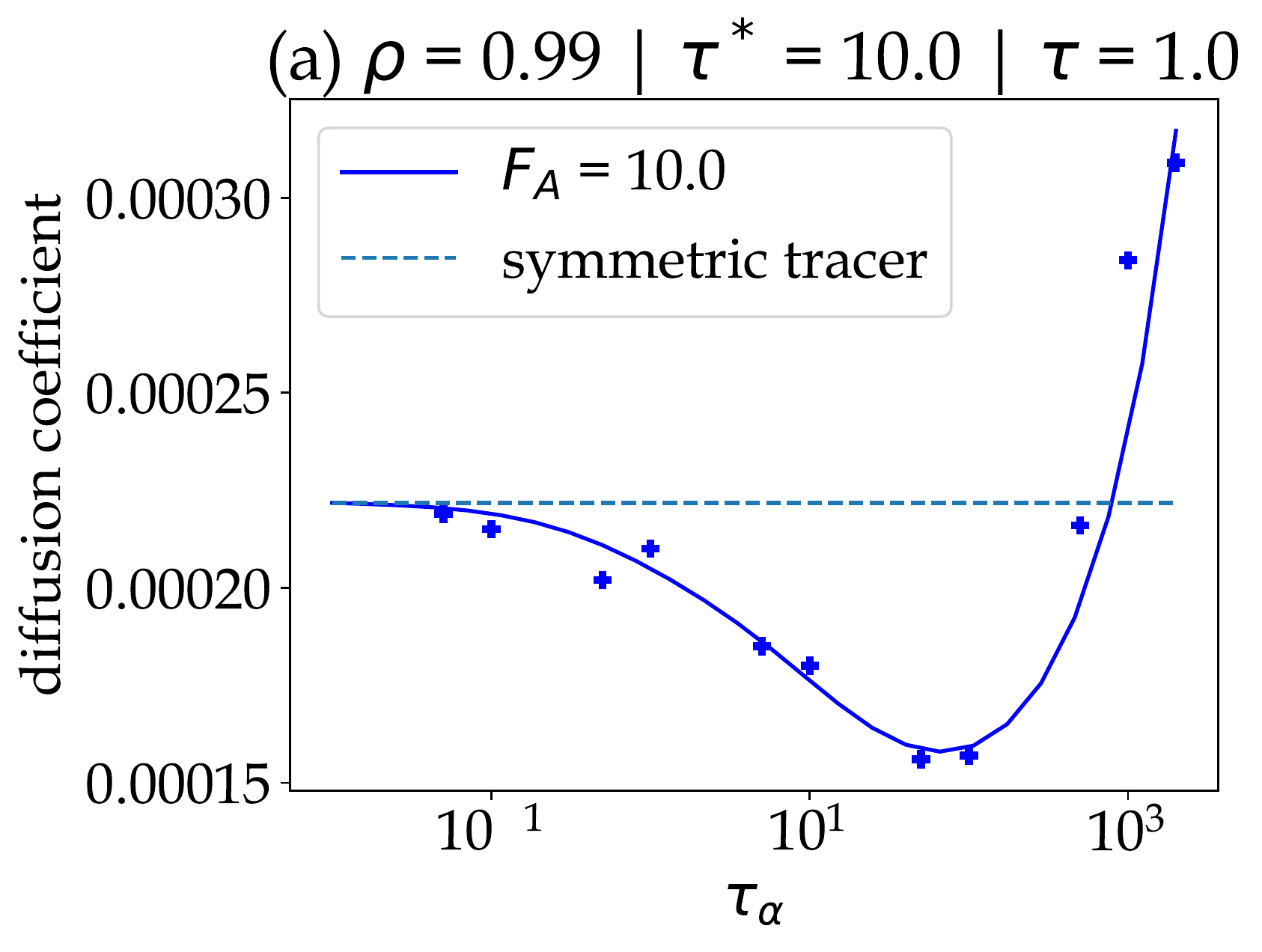}
\caption{Non-monotony of $D$ as a function of the persistence time $\tau_\alpha$, at density $\rho=0.1$ (a) and $\rho=0.99$ (b). 
}
\label{fig:non-montony_tau_alpha}
\end{figure}

\emph{Non-monotony on the parameters controlling activity.---} We now study the dependence of the diffusion coefficient on the persistence time $\tau_\alpha$. The asymptotic limits $\tau_\alpha\to 0$ and $\tau_\alpha\to \infty$ are known: when the persistence time becomes very small, the diffusion coefficient is finite and equal to that of a passive tracer \cite{Nakazato1980}, while in the limit of an infinitely persistent tracer, the diffusion coefficient is expected to diverge (except in the specific limit of fixed obstacles $\tau^*\to\infty$). Our analysis reveals that the diffusion coefficient can exhibit a nonmonotonic behavior between these two limits, as previously observed in the low-density limit \cite{Bertrand2018a}.
This effect remains when $\tau^*/\tau <\infty$, but was only studied in the situation of an infinite active force, i.e. in the limit where the tracer cannot step sideways from its persistence direction \cite{Bertrand2018a}. Here, we go one step further and study the effect of the active force {for an arbitrary density of crowders on the lattice}. For a given value of $\rho$ and $\tau^*/\tau$, the non-monotony of the diffusion coefficient persists as long as the active force is large enough, as shown in Fig. \ref{fig:non-montony_tau_alpha}. {This effect results from the competition between the different timescales governing the diffusion of the tracer, and can be captured with simple analytical arguments. { A phase diagram, which represents the critical value of $\tau^*/\tau $ above which $D$ becomes a non-monotonic function of $\tau_\alpha$ (for given density $\rho$ and force $F_A$) is given in SM (Fig. S2) \cite{SM}.}}
{We also observe a non-monotony of the diffusion coefficient with the active force $F_A$, which is reminiscent of previous observations in the case of an infinitely persistent tracer \cite{Illien2017c}.}

 \emph{Low- and high-density regimes.---} Finally, as shown in Fig. \ref{fig:drho_2D}, the decoupling approximation is accurate for the whole range of density $0 \leq \rho \leq 1$. In addition, we argue that it becomes exact in the low- and high-density regimes, that we explore here. This claim relies on the exactness of (i) the theory of Nakazato-Kitahara in the case of symmetric passive tracer \cite{Nakazato1980}; and (ii) the microscopic theory of a driven passive tracer in these limits~\cite{Illien2017c}.

{ We expand the density profiles $h_\mu^{(\chi)}$ and the correlation functions $\gt_\mu^{(\chi)}$ in the limits of $\rho\to0$ and $\rho\to 1$. In these limits, the diffusion coefficient of the tracer is expanded as $D = D_0 + \rho \mathcal{D}_0 + \mathcal{O}(\rho^2)$ (resp. $D=(1-\rho) \mathcal{D}_1 + \mathcal{O}[(1-\rho)^2]$) where the expressions of $D_0$ and $\mathcal{D}_0$ (resp. $\mathcal{D}_1$) are expressed in terms of the leading order expressions of the coefficients $h_\mu^{(\chi)}$ and $\gt_\mu^{(\chi)}$ in the low-density (resp. high-density) limit. The latter are found to be solutions of linear systems. These solutions, together with the expansions of $D$, yields an  \emph{explicit} expression of the diffusion coefficient of the tracer in terms of all the parameters of the problem in these regimes  (see Sections IX and XI of SM and Fig. \ref{fig:drho_2D}).}

In the low-density limit, this result is the continuous-time counterpart of previous low-density approaches, which relied on a specific dynamics. A comparison between the results from our decoupling approximation and the results from Ref. \cite{Bertrand2018a} is given in SM \cite{SM}. Since the two dynamics are different, the two calculations of the diffusion coefficient do not match quantitatively, but display a good qualitative agreement. 

These expansions give fully explicit expressions of the diffusion coefficient both in the low- and high-density regimes, which we furthermore argue to be exact. {Indeed, in both  limits of a driven tracer ($\tau_\alpha\to\infty$) and of a passive tracer ($\tau_\alpha\to 0$), a similar decoupling approximation was compared to exact approaches that focused: (i) on the low-density limit, in which the diffusion of the tracer is seen as a succession of scattering events due to interactions with independent obstacles (at leading order in $\rho$) \cite{Leitmann2013, Leitmann2017}; (ii) on the high-density limit, in which the diffusion of the tracer is mediated by the diffusion of vacancies, which explore the lattice independently (at  leading order in $1-\rho$)  \cite{Brummelhuis1989a,Benichou2002a,Illien2013a,Benichou2013c,Illien2014}. This, together with the very good agreement between the decoupling approximation and numerical results, points towards the exactness of the present approximation. Showing such exactness would require to obtain exact results for the diffusion of an active tracer using the methods mentioned above, and this will be investigated in future work.}

{We hope that the present approach will allow to establish connections with recent experimental observations on living organisms \cite{Licata2016,Croze2011} or self-propelled particles \cite{Lozano2019} in crowded environments. Moreover, it will be technically challenging but particularly interesting to study the opposite situation of a passive tracer in an dense active environment -- a situation that has recently been the object of theoretical approaches \cite{Reichert2020}.}

\begin{acknowledgments}
AS acknowledges partial support from MIUR project PRIN201798CZLJ and from Program (VAnviteLli pEr la RicErca: VALERE) 2019 financed by the University of Campania “L. Vanvitelli”.
\end{acknowledgments}


%

\end{document}


\title{Microscopic theory for the diffusion of an active particle in a crowded environment\\ \emph{Supplemental Material}}

\author{Pierre Rizkallah}
\affiliation{Sorbonne Universit\'e, CNRS, Laboratoire de Physico-Chimie des \'Electrolytes et Nanosyst\`emes Interfaciaux (PHENIX), 4 Place Jussieu, 75005 Paris, France}

\author{Alessandro Sarracino}
\affiliation{Dipartimento di Ingegneria, Universit\`a della Campania ”Luigi Vanvitelli”, 81031 Aversa (CE), Italy}

\author{Olivier Bénichou}
\affiliation{Sorbonne Universit\'e, CNRS, Laboratoire de Physique Th\'eorique de la Mati\`ere Condens\'ee (LPTMC), 4 Place Jussieu, 75005 Paris, France}

\author{Pierre Illien}
\affiliation{Sorbonne Universit\'e, CNRS, Laboratoire de Physico-Chimie des \'Electrolytes et Nanosyst\`emes Interfaciaux (PHENIX), 4 Place Jussieu, 75005 Paris, France}

\setcounter{equation}{0}
\setcounter{figure}{0}

\renewcommand{\theequation}{S\arabic{equation}}
\renewcommand{\thefigure}{S\arabic{figure}}

\renewcommand*{\citenumfont}[1]{S#1}
\renewcommand*{\bibnumfmt}[1]{[S#1]}

\onecolumngrid

%
%
%
%

\maketitle

\onecolumngrid

\tableofcontents

\section{Definition of the operator $\mathcal{L}_\chi$}

If we denote by $\eta^{{\rr},\mu}$ the configuration obtained from configuration $\eta$ by switching the values of $\eta_{\rr}$ and $\eta_{\rr + \ee_\mu}$, the evolution operator in state $\chi$ is defined by : 
\begin{eqnarray}
\mathcal{L}_\chi P_\chi=&&
  \sum_{\mu=1}^d\sum_{{\rr}\neq\RR-\ee_\mu,\RR} \left[ P_\chi(\RR,\eta^{{\rr},\mu};t)-P_\chi(\RR,\eta;t)\right]\nonumber\\
&&+\frac{2d\tau^*}{\tau}\sum_{\mu} p_\mu^{(\chi)} 
[\left(1-\eta_{\RR} \right)P_\chi(\RR-\ee_{\mu},\eta;t) -\left(1-\eta_{\RR+\ee_{\mu}}\right)P_\chi(\RR,\eta;t)]
\end{eqnarray}
 {The first term of the operator corresponds to the diffusion of bath particles, the second term corresponds to the diffusion of the tracer}.



\section{Expression of the operator $\mathcal{G}^{(\chi)}$}

We give here the expression of the operator $\mathcal{G}^{(\chi)}$ involved in Eq. (4) from the main text:
\begin{align}
&\mathcal{G}^{(\chi)}h_{\rr}^{(\chi)} \equiv   \dfrac{2d\tau^*}{\tau} \textstyle\sum_{\mu = \pm 1} \mu  p_\mu^{(\chi)} \nabla_\mu h_{\rr}^{(\chi)} \left(1-\rho-h_{\ee_{\mu}}^{(\chi)}\right)  \nonumber\\
&- \dfrac{2d\tau^*}{\tau} \textstyle \sum_\mu p_\mu^{(\chi)} \gt_{\ee_{\mu}}^{(\chi)}\nabla_\mu h_{\rr}^{(\chi)} \nonumber\\
&  +  \dfrac{\alpha}{2d-1} \textstyle\sum_{\chi'\neq\chi} \left(X^\infty_{\chi'} -X^\infty_\chi\right)\left(h_{\rr}^{(\chi')} - h_{\rr}^{(\chi)} \right),
\end{align}
where $X^\infty_\chi=\lim_{t\to\infty}\moy{X_t}_\chi$ is the average of the position of the tracer conditioned on state $\chi$ in the stationary limit (see below).

\section{Determination of the average position of the tracer $X^\infty_\chi$}

When $\alpha \neq 0$, the mean position of the tracer in a given state $\moy{X_t}_\chi$ tends to a finite limit $X_\chi^\infty = \lim_{t\to\infty} \moy{X_t}_\chi$.

To compute these quantities, we consider the evolution equations derived from the master equation : 
\begin{equation}
 \dfrac{\dd}{\dd t}\moy{X_t}_\chi = \frac{1 \sigmaa}{\tau} \left\lbrace p_1^{(\chi)} \left[1-k_{\ee_1}^{(\chi)}\right]-p_{-1} ^{(\chi)}\left[1-k_{\ee_{-1}}^{(\chi)}\right]  \right\rbrace + \dfrac{\alpha}{2d\tau^*} \left(-\moy{X_t}_\chi + \dfrac{1}{2d-1} \sum_{\chi'\neq\chi} \moy{X_t}_{\chi'}\right)
\end{equation}
{The last term accounts for the random reorientations of the tracer.} In the stationary regime, the right hand side vanishes. This gives a linear system of equations, which solutions are $X^\infty_\chi = 0$ for $\chi \neq \pm 1$ and :
\begin{equation}
    \label{eq:Xinf}
    X^\infty_{1} = -X^\infty_{-1} = \frac{\sigmaa (2d-1) \tau^*}{\tau \alpha} \left\lbrace p_1^{(\chi)} \left[1-k_{\ee_1}^{(\chi)}\right]-p_{-1} ^{(\chi)}\left[1-k_{\ee_{-1}}^{(\chi)}\right]  \right\rbrace
\end{equation}

{

\section{Expression of the diffusion coefficient}
To compute the evolution of observables from the master equation, we multiply it by the observable, then we sum on all possible configurations. Here we denote $R_1 = \RR \cdot \ee_1$. We first compute the equation on the mean square displacement conditioned to the active force being in direction $\chi$ : 

\begin{align}
2d\frac{\dd}{\dd t} \sum_{\RR, \eta} {R_1}^2 P_\chi(\RR,\eta;t) = & 2d\sum_{\RR, \eta}\left\lbrace \sum_{\mu=1}^d\sum_{{\rr}\neq\RR-\ee_\mu,\RR} R_1^2\left[ P_\chi(\RR,\eta^{{\rr},\mu};t)-P_\chi(\RR,\eta;t)\right] \right.\nonumber\\
&+\frac{1}{\tau}\sum_{\mu} p_\mu^{(\chi)}R_1^2  
[\left(1-\eta_{\RR} \right)P_\chi(\RR-\ee_{\mu},\eta;t) -\left(1-\eta_{\RR+\ee_{\mu}}\right)P_\chi(\RR,\eta;t)]\nonumber\\
&\left.- \alpha R_1^2 P_\chi(\RR,\eta;t) +\frac{\alpha}{2d-1} \sum_{\chi'\neq\chi} R_1^2 P_{\chi'}(\RR,\eta;t)\right\rbrace \nonumber\\
\frac{\dd}{\dd t} \moy{{X_t}^2}_\chi = & 0 + \frac{1}{\tau}\sum_{\mu} p_\mu^{(\chi)} 
\left[\moy{\left(1-\eta_{\XX_t+\ee_\mu} \right)(X_t^2 + 2 X_t (\ee_1 \cdot \ee_\mu) + (\ee_1 \cdot \ee_\mu))^2}_\chi
-\moy{X_t^2\left(1-\eta_{\RR+\ee_{\mu}}\right)}_\chi\right] \nonumber\\
& - \moy{X_t^2}_\chi +\frac{\alpha}{2d-1} \sum_{\chi'\neq\chi} \moy{X_t^2}_{\chi'} \nonumber\\
= & \frac{1}{\tau} \left\lbrace p_1^{(\chi)} \left[1-k_{\ee_1}^{(\chi)}(t)\right]+p_{-1} ^{(\chi)}\left[1-k_{\ee_{-1}}^{(\chi)}(t)\right] \right\rbrace \nonumber\\
 & + \frac{2}{\tau}\left[p_1^{(\chi)} \left( \moy{X_t}_\chi - g_{\ee_1}^{(\chi)} \right)-  p_{-1}^{(\chi)} \left( \moy{X_t}_\chi - g_{\ee_{-1}}^{(\chi)} \right) \right] - \moy{X_t^2}_\chi +\frac{\alpha}{2d-1} \sum_{\chi'\neq\chi} \moy{X_t^2}_{\chi'}
\end{align}
We then get for the long time limit of the derivative of the mean square displacement (the mean position is always 0): 
\begin{align}
2D = \frac{\dd}{\dd t} \moy{{X_t}^2}  = & \frac{1}{2d}\sum_{\chi} \frac{\dd}{\dd t} \moy{{X_t}^2}_\chi  \nonumber\\
  = &  \dfrac{1}{2d}\sum_{\chi}  \frac{1}{\tau} \left\lbrace p_1^{(\chi)} \left[1-k_{\ee_1}^{(\chi)}(t)\right]+p_{-1} ^{(\chi)}\left[1-k_{\ee_{-1}}^{(\chi)}(t)\right] \right\rbrace  \nonumber\\
 & + \dfrac{1}{2d}\sum_{\chi}  \frac{2}{\tau}\left[p_1^{(\chi)} \left( \moy{X_t}_\chi - g_{\ee_1}^{(\chi)} \right)-  p_{-1}^{(\chi)} \left( \moy{X_t}_\chi - g_{\ee_{-1}}^{(\chi)} \right) \right] \nonumber\\
= & \dfrac{1}{2d}\sum_{\chi}\frac{1}{\tau} \left\lbrace p_1^{(\chi)} \left[1-k_{\ee_1}^{(\chi)}(t)\right]+p_{-1} ^{(\chi)}\left[1-k_{\ee_{-1}}^{(\chi)}(t)\right] \right\rbrace - \frac{2}{\tau}\left[p_1^{(\chi)} \gt_{\ee_1} -  p_{-1}^{(\chi)} \gt_{\ee_{-1}} \right] \nonumber\\
 & + \dfrac{1}{2d} \sum_{\chi} \frac{2X_\chi^\infty}{\tau}\left[p_1^{(\chi)} \left( 1 - k_{\ee_1}^{(\chi)} \right)-  p_{-1}^{(\chi)} \left(1 - k_{\ee_{-1}}^{(\chi)} \right) \right]
\end{align}
Together with \eqref{eq:Xinf}, this gives the expression given in main text.

}




\section{Result by Nakazato and Kitahara}
In the limit of a symmetric tracer ($F_A = 0$ or $\alpha \to \infty$), we retrieve the result by Nakazato and Kitahara \cite{Nakazato1980}, which reads: 
\begin{equation}
D = \dfrac{1}{2d\tau}(1-\rho)\left[1- \dfrac{2\rho\frac{\tau^*}{\tau}\gamma}{2d\left[1+\frac{\tau^*}{\tau}(1-\rho)\right] - \left[1+\frac{\tau^*}{\tau}(1-3\rho)\gamma\right]} \right],
\end{equation}
where $\gamma$ is a geometry-dependent constant:
\begin{equation}
    \gamma = 
    \begin{dcases}
    4-\frac{8}{\pi} & \text{for a 2D infinite lattice,} \\
    \frac{4}{(2\pi)L}\sum_{k=0}^{L-1}\int_0^{2\pi}\frac{\sin(q)^2}{2-\cos(q)-\cos\left(\frac{2\pi k}{L}\right)}\dd q &   \text{for a 2D stripe-like lattice,} \\
    \frac{6}{(2\pi)^3}\int_0^{2\pi}\int_0^{2\pi}\int_0^{2\pi} \frac{\sin(q_1)^2}{3-\cos(q_1)-\cos(q_2)-\cos(q_3)}\dd q_1 \dd q_2 \dd q_3 & \text{for a 3D infinite lattice.} 
    \end{dcases}
\end{equation}


\section{Analytical resolution}

We begin by computing the Fourier transforms of Eqs. (3) and (4) from the main text :
\begin{align}
2d\tau^* \partial_t H^{(\chi)}(\qq;t) = & K^{(\chi)}H^{(\chi)}+K_0^{(\chi)}-\alpha H^{(\chi)}+\frac{\alpha}{2d-1}\sum_{\chi'\neq\chi}H^{(\chi')} \label{eq:FourierH}\\
2d\tau^* \partial_t G^{(\chi)}(\qq;t) =& K^{(\chi)}G^{(\chi)}+ J_H^{(\chi)}H^{(\chi)}+J_0^{(\chi)}  + \dfrac{\alpha}{2d-1} \sum_{\chi'\neq\chi} \left(\moy{X_t}_{\chi'} - \moy{X_t}_{\chi} \right)\left(H^{(\chi')} - H^{(\chi)} \right) \nonumber\\
&  -\alpha G^{(\chi)}+\frac{\alpha}{2d-1}\sum_{\chi'\neq\chi}G^{(\chi')}  \label{eq:FourierG}
\end{align}
If we denote $h_\mu^{(\chi)} = h_{\ee_\mu}^{(\chi)}$ and same convention for $\gt^{(\chi)}$, the functions involved in the Fourier transform are (remember $q_\mu = \qq\cdot\ee_\mu$ and $\ee_{-\mu} = - \ee_\mu$):
\begin{eqnarray}
K^{(\chi)}(\qq;t) & = & \sum_{\mu}     \left(\ex{-\ii q_\mu} -1\right)A_\mu^{(\chi)}(t)\\
K_0^{(\chi)}(\qq;t) & = &  \sum_{\mu}     \left(\ex{\ii q_\mu} -1\right)A_\mu^{(\chi)}(t) h_{\mu}^{(\chi)}(t) + 2\ii\sum_{j=1}^d \left[A_j^{(\chi)} -A_{-j}^{(\chi)} \right]\sin ( q_j) \\
J_H^{(\chi)}(\qq;t) & = & \sum_{\mu=\pm 1} \mu\left(A_\mu^{(\chi)} - 1 \right) \left(\ex{-\ii q_\mu} - 1\right)  -\frac{2d\tau^*}{\tau} \sum_{\mu} p^{(\chi)}_\mu \gt^{(\chi)}_\mu  (\ex{-\ii q_\mu}-1) 
\\
J_0^{(\chi)}(\qq;t) & = & \sum_\mu (\ex{\ii q_\mu}-1) \left(A^{(\chi)}_\mu-\frac{2d\tau^*}{\tau}p^{(\chi)}_\mu h^{(\chi)}_\mu \right) \gt^{(\chi)}_\mu - \rho \frac{2d\tau^*}{\tau} 2\ii  \sum_{j=1}^d \sin (q_j) (p^{(\chi)}_j \gt^{(\chi)}_j - p^{(\chi)}_{-j} \gt^{(\chi)}_{-j}) \nonumber \\
&& -\sum_{\mu=\pm 1} \mu (A^{(\chi)}_\mu-1)\left(\rho\ex{-\ii q_\mu} + h_\mu^{(\chi)}\right)
\end{eqnarray} 

Theses equations can be rewritten in a more convenient form by the following method.
First, symmetry gives the following relations: \\
For any $\mu$ and $\chi \notin \left\lbrace\mu , -\mu\right\rbrace$ :
\begin{align}
h_\mu^{(\mu)} & = h_1^{(1)} \\
h_\mu^{(-\mu)} & = h_{-1}^{(1)} \\
 h_\mu^{(\chi)} & = h_2^{(1)} 
\end{align}
and for $\mu, \chi \notin \left\lbrace -1, 1 \right\rbrace$ : 
\begin{align}
\gt_{-1}^{(-1)} & = -\gt_{1}^{(1)} \\
\gt_{1}^{(-1)} & = -\gt_{-1}^{(1)} \\
\gt_{\mu}^{(-1)} & = -\gt_\mu^{(1)} = -g_{2}^{(1)} \\
\gt_{1}^{(-\chi)} & = -\gt_{-1}^{(-\chi)} = \gt_{1}^{(\chi)} = -\gt_{- 1}^{(\chi)}\\
\gt_{\pm \mu}^{( \pm \chi)} & = 0 
\end{align}

Thanks to these relations, we have only 3 independent values for $h_\mu^{(\chi)}$ and 4 for $\gt_\mu^{(\chi)}$. Therefore, the following vectors contain all the values of interest: 
\begin{equation}
\boldsymbol{h} = \left(\begin{matrix}
h_1^{(1)} \\
h_1^{(-1)}\\
h_1^{(2)} \\
h_1^{(-2)}\\
\end{matrix}\right)
\text{~~~and ~~~} \boldsymbol{g} = \left(\begin{matrix}
\gt_1^{(1)}\\
\gt_{-1}^{(1)} \\
\gt_2^{(1)} \\
\gt_1^{(2)} \\
\end{matrix}\right)
\end{equation}
Using these symmetries, the matricial equation on $\boldsymbol{H} = \left( H^{(\chi)}(\qq) \right)_{\chi \in \lbrace \pm 1, \pm 2 \rbrace}$ can be written in the following form (to lighten notations, we don't write dependencies in $\qq$):
\begin{equation}
\boldsymbol{M}\cdot \boldsymbol{H} + \boldsymbol{\Lambda} \cdot \boldsymbol{h} + \boldsymbol{S} = 0
\end{equation}

The different matrices are precised below. Finally we can perform the inverse Fourier transform and get the nonlinear system on $h_\mu^{(\chi)}$ (we recall that they also appear in the definition of $\boldsymbol{M}$ through $A_\mu^{(\chi)}$): 

\begin{equation}
\label{eqh}
\left(\boldsymbol{1} + \int_{\qq} \ex{-\ii\sigmaa q_1} \boldsymbol{M}^{-1}\boldsymbol{\Lambda} \right)\cdot \boldsymbol{h} = -\int_{\qq} \ex{-\ii\sigmaa q_1} \boldsymbol{M}^{-1} \boldsymbol{S},
\end{equation}
where we recall the shorthand notation $\int_{\qq} = \int_{[-\pi,\pi]^d} \frac{\dd\qq}{(2\pi)^d}$ for the inverse Fourier transform. This nonlinear system can be solved numerically. Similarly, the matricial equation on $\boldsymbol{G} = \left( G^{(\chi)}(\qq) \right)_{\chi \in \lbrace \pm 1, \pm 2 \rbrace}$ can be written with matrices detailed below : 
\begin{equation}
\boldsymbol{M} \cdot \boldsymbol{G} + \boldsymbol{\Lambda_G} \cdot \boldsymbol{g} + \boldsymbol{X}\cdot \boldsymbol{H} + \boldsymbol{E} + \boldsymbol{F} = 0
\label{eqMG}
\end{equation}
We can perform an inverse Fourier transform on this system, to get the final linear system on $\gt_\mu^{(\chi)}$ :
\begin{equation}
\label{eqg}
(\boldsymbol{1}+\boldsymbol{L})\cdot \boldsymbol{g} = -\boldsymbol{B}
\end{equation}
The matrices involved in the equations above are directly obtained from the Fourier transforms \eqref{eq:FourierH} and \eqref{eq:FourierG}. For the equation on the profiles :
\begin{equation}
\boldsymbol{\Lambda} = \left(\begin{matrix} 
\left(\ex{\ii\sigmaa q_1} -1\right)A_1^{(1)} & \left(\ex{-\ii\sigmaa q_1} -1\right)A_{-1}^{(1)} &
2\left(\cos\left(\sigmaa q_2\right) -1\right)A_2^{(1)} & 0 \\

\left(\ex{-\ii\sigmaa q_1} -1\right)A_1^{(1)} & \left(\ex{\ii\sigmaa q_1} -1\right)A_{-1}^{(1)} &
2\left(\cos\left(\sigmaa q_2\right) -1\right)A_2^{(1)} & 0\\

\left(\ex{\ii\sigmaa q_2} -1\right)A_1^{(1)} & \left(\ex{-\ii\sigmaa q_2} -1\right)A_{-1}^{(1)} &
2\left(\cos\left(\sigmaa q_1\right) -1\right)A_2^{(1)} & 0\\

\left(\ex{-\ii\sigmaa q_2} -1\right)A_1^{(1)} & \left(\ex{\ii\sigmaa q_2} -1\right)A_{-1}^{(1)} &
2\left(\cos\left(\sigmaa q_1\right) -1\right)A_2^{(1)} & 0
 
\end{matrix}\right)
\end{equation}
and: 
\begin{equation}
\boldsymbol{S} = 2\ii \rho \dfrac{2d\tau^*}{\tau} \left[p_1^{(1)}(1-\rho - h_1^{(1)}) -p_{-1}^{(1)}(1- \rho - h_{1}^{(-1)}) \right]\left(\begin{matrix}
 \sin (\sigmaa q_1)\\
-\sin (\sigmaa q_1)\\
 \sin (\sigmaa q_2)\\
-\sin (\sigmaa q_2)
\end{matrix}\right)
\end{equation}
In the same way, the matrices involved in the equation for the correlation between position and occupation are : 
\begin{equation}
\boldsymbol{E} = \sigmaa\left(\begin{matrix}
H^{(1)}\left((A_1^{(1)}-1) (\ex{-\ii \sigmaa	q_1}-1) - (A_{-1}^{(1)}-1) (\ex{\ii \sigmaa	q_1} - 1)\right)\\
H^{(-1)}\left((A_{-1}^{(1)}-1) (\ex{-\ii \sigmaa q_1} - 1) - (A_1^{(1)}-1) ( \ex{\ii \sigmaa	q_1} -1)\right)\\
-2\ii (A_2^{(1)} - 1) H^{(2)} \sin(\sigmaa	q_1) \\
-2\ii (A_2^{(1)} - 1) H^{(-2)} \sin(\sigmaa q_1) \\
\end{matrix}\right)
\end{equation}
\begin{equation}
\boldsymbol{X} = 
\dfrac{\alpha}{2d-1}\left(\begin{matrix} 
X_{\chi\chi} = -\sum_{\chi'\neq\chi} \left(\moy{X}_{\chi'} - \moy{X}_{\chi} \right) \\
X_{\chi\chi'} = \moy{X}_{\chi'} - \moy{X}_{\chi}
\end{matrix}\right)_{\chi\chi'}
\end{equation}

\begin{align}
\boldsymbol{\Lambda_G}  =& -\frac{2d\tau^*}{\tau}\left(\begin{matrix}
H^{(1)}p^{(1)}_1(\ex{-\ii\sigmaa q_1}-1) & H^{(1)}p^{(1)}_{-1}(\ex{\ii\sigmaa q_1}-1) & 2H^{(1)}p^{(1)}_2(\cos(\sigmaa q_2)-1) & 0\\
-H^{(-1)}p^{(-1)}_{-1}(\ex{\ii\sigmaa q_1}-1) & -H^{(-1)}p^{(-1)}_{1}(\ex{-\ii\sigmaa q_1}-1) & -2H^{(-1)}p^{(-1)}_2(\cos(\sigmaa q_2)-1) & 0\\
0 & 0 & 0 & 2H^{(2)}p^{(2)}_{1}(\cos(\sigmaa q_1)-1)\\
0 & 0 & 0 & 2H^{(-2)}p^{(-2)}_{1}(\cos(\sigmaa q_1)-1)
\end{matrix}\right) \nonumber\\
& - \rho \frac{2d\tau^*}{\tau} 2\ii \left(\begin{matrix}
\sin(\sigmaa q_1)p^{(1)}_1 & -\sin(\sigmaa q_1)p^{(1)}_{-1} & 0 & 0\\
\sin(\sigmaa q_1)p^{(-1)}_{-1} & -\sin(\sigmaa q_1)p^{(-1)}_{1} & 0 & 0\\
0&0&0&2 \sin(\sigmaa q_1)p^{(2)}_{1}\\
0&0&0&2 \sin(\sigmaa q_1)p^{(-2)}_{1}
\end{matrix}\right) \nonumber\\
& + \left(\begin{matrix}
\left(\ex{\ii\sigmaa q_1} -1\right)A_1^{(1)} & \left(\ex{-\ii\sigmaa q_1} -1\right)A_{-1}^{(1)} & 2\left(\cos\left(\sigmaa q_2\right) -1\right)A_2^{(1)} & 0 \\
-\left(\ex{-\ii\sigmaa q_1} -1\right)A_1^{(1)} & -\left(\ex{\ii\sigmaa q_1} -1\right)A_{-1}^{(1)} & -2\left(\cos\left(\sigmaa q_2\right) -1\right)A_2^{(1)} & 0 \\
0 & 0 & 0 & 2\ii A_2^{(1)} \sin(\sigmaa q_1)  \\
0 & 0 & 0 & 2\ii A_2^{(1)} \sin(\sigmaa q_1) 
\end{matrix}\right) \nonumber\\
& - \dfrac{2d\tau^*}{\tau}\left(\begin{matrix}
\left(\ex{\ii\sigmaa q_1} -1\right)p_1^{(1)}h_1^{(1)} & \left(\ex{-\ii\sigmaa q_1} -1\right)p_1^{(-1)}h_1^{(-1)} & 2\left(\cos\left(\sigmaa q_2\right) -1\right)p_1^{(2)}h_1^{(2)} & 0 \\
-\left(\ex{-\ii\sigmaa q_1} -1\right)p_1^{(1)}h_1^{(1)} & -\left(\ex{\ii\sigmaa q_1} -1\right)p_1^{(-1)}h_1^{(-1)} & -2\left(\cos\left(\sigmaa q_2\right) -1\right)p_1^{(2)}h_1^{(2)} & 0 \\
0 & 0 & 0 & 2p_1^{(2)}h_1^{(2)}\left(\cos(\sigmaa q_1) -1\right) \\
0 & 0 & 0 & 2p_1^{(2)}h_1^{(2)}\left(\cos(\sigmaa q_1) -1\right)
\end{matrix}\right)\\
\boldsymbol{F} = &-\sigmaa\left(\begin{matrix}
(A_1^{(1)}-1)\left(\rho\ex{-\ii\sigmaa q_1} + h_1^{(1)}\right) - (A_{-1}^{(1)}-1)\left(\rho\ex{\ii\sigmaa q_1} + h_{-1}^{(1)}\right) \\
(A_{-1}^{(1)}-1)\left(\rho\ex{-\ii\sigmaa q_1} + h_1^{(-1)}\right) - (A_{1}^{(1)}-1)\left(\rho\ex{\ii\sigmaa q_1} + h_{-1}^{(-1)}\right) \\
(A_2^{(1)}-1)\left(\rho\ex{-\ii\sigmaa q_1} + h_1^{(2)}\right) - (A_{2}^{(1)}-1)\left(\rho\ex{\ii\sigmaa q_1} + h_{-1}^{(2)}\right) \\
(A_2^{(1)}-1)\left(\rho\ex{-\ii\sigmaa q_1} + h_1^{(-2)}\right) - (A_{2}^{(1)}-1)\left(\rho\ex{\ii\sigmaa q_1} + h_{-1}^{(-2)}\right) 
\end{matrix}\right)
\end{align}
Then we perform the inverse Fourier transforms (where $[ \cdot ]_i$ means that we take the $i$-th line) : 
\begin{equation}
\boldsymbol{L} = \left(\begin{matrix}
\int_{\qq} \ex{-\ii\sigmaa q_1} \left[\boldsymbol{M}^{-1}\boldsymbol{\Lambda_G}\right]_1 \\
\int_{\qq} \ex{\ii\sigmaa q_1} \left[\boldsymbol{M}^{-1}\boldsymbol{\Lambda_G}\right]_1 \\
\int_{\qq} \ex{-\ii\sigmaa q_2} \left[\boldsymbol{M}^{-1}\boldsymbol{\Lambda_G}\right]_1 \\
\int_{\qq} \ex{-\ii\sigmaa q_1} \left[\boldsymbol{M}^{-1}\boldsymbol{\Lambda_G}\right]_3 
\end{matrix}\right)
\text{~~ and ~~} \boldsymbol{B} =  \left(\begin{matrix}
\int_{\qq} \ex{-\ii\sigmaa q_1} \left[\boldsymbol{M}^{-1}(\boldsymbol{X}\cdot \boldsymbol{H} + \boldsymbol{E} + \boldsymbol{F})\right]_1 \\
\int_{\qq} \ex{\ii\sigmaa q_1} \left[\boldsymbol{M}^{-1}(\boldsymbol{X}\cdot \boldsymbol{H} + \boldsymbol{E} + \boldsymbol{F})\right]_1 \\
\int_{\qq} \ex{-\ii\sigmaa q_2} \left[\boldsymbol{M}^{-1}(\boldsymbol{X}\cdot \boldsymbol{H} + \boldsymbol{E} + \boldsymbol{F})\right]_1 \\
\int_{\qq} \ex{-\ii\sigmaa q_1} \left[\boldsymbol{M}^{-1}(\boldsymbol{X}\cdot \boldsymbol{H} + \boldsymbol{E} + \boldsymbol{F})\right]_3 
\end{matrix}\right)
\end{equation}





\section{Non-monotony of the diffusion coefficient as a function of the active force}

\begin{figure}
\includegraphics[width=0.6\columnwidth]{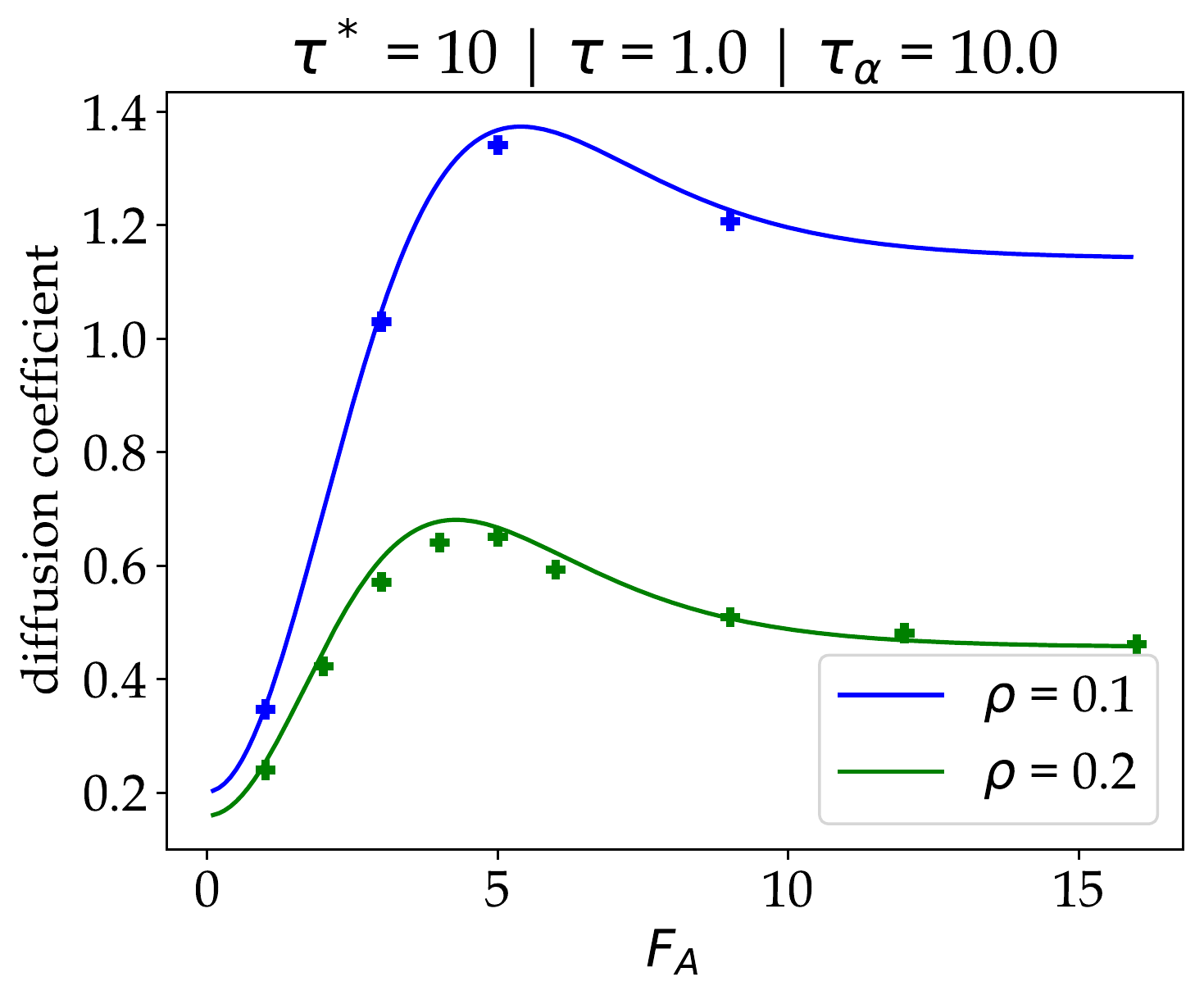}
\caption{Non-monotony of the diffusion coefficient as a function of the magnitude of the active force $F_A$. The symbols represent the results from Monte Carlo simulations, the solid lines are the results from our  analytical approach.}
\label{fig:non-monotony_F}
\end{figure}

 The diffusion coefficient of the tracer is plotted as a function of the active force $F_A$ on Fig. \ref{fig:non-monotony_F}.

\section{Simple arguments for the non-monotony of $D$}


We can recover most of the qualitative results found at low density and large force thanks to simple physical considerations, which extend the arguments that were put forward in the case of a passive driven tracer \cite{Benichou2014}. At low density, the obstacles are independent and the probability for the tracer to find an obstacle at a given site is $\rho$. For a large force $F_A\to\infty$, we can neglect the probability that the tracer moves in the direction opposite to the force. 

We can therefore approximate the characteristic time between two jumps of the tracer by the sum of the internal jump time $\tau$ and the mean trapping time which can be evaluated by considering that the escape from a trap is caused by three independent events following exponential laws: (i) the obstacle moves in a transverse direction with characteristic time $\frac{2d\tau^*}{(2d-2)}$; (ii) the active force changes direction with characteristic time $\tau_\alpha$; (iii) the tracer moves in a direction transverse to the active force with characteristic time $\tau/(1-p_\chi^{(\chi)}-p_{-\chi}^{(\chi)})$. Consequently the mean trapping time follows an exponential law of characteristic time $\tau_p$ given by
\begin{equation}
    \frac{1}{\tau_p} = \frac{(2d-2)}{2d\tau^*} + \dfrac{1}{\tau_\alpha} + \frac{(1-p_1^{(1)} - p_{-1}^{(1)})}{\tau} 
\end{equation}
Under these hypotheses, the tracer behaves like an isolated persistent tracer (absence of obstacles) with characteristic jump time equal to $\tau + \rho \tau_p$. Equation (2) from the main text then gives the following formula for the diffusion coefficient : 
\begin{equation}
    D = \frac{1}{2d(\tau + \rho \tau_p)} \left[1  + \frac{(2d-1)\tau_\alpha}{d(\tau + \rho \tau_p)}\left(p_1^{(1)} - p_{-1}^{(1)}\right)^2\right] 
    \label{eq:D_simple}
\end{equation}

\begin{figure}
\begin{center}
\includegraphics[scale = 0.75]{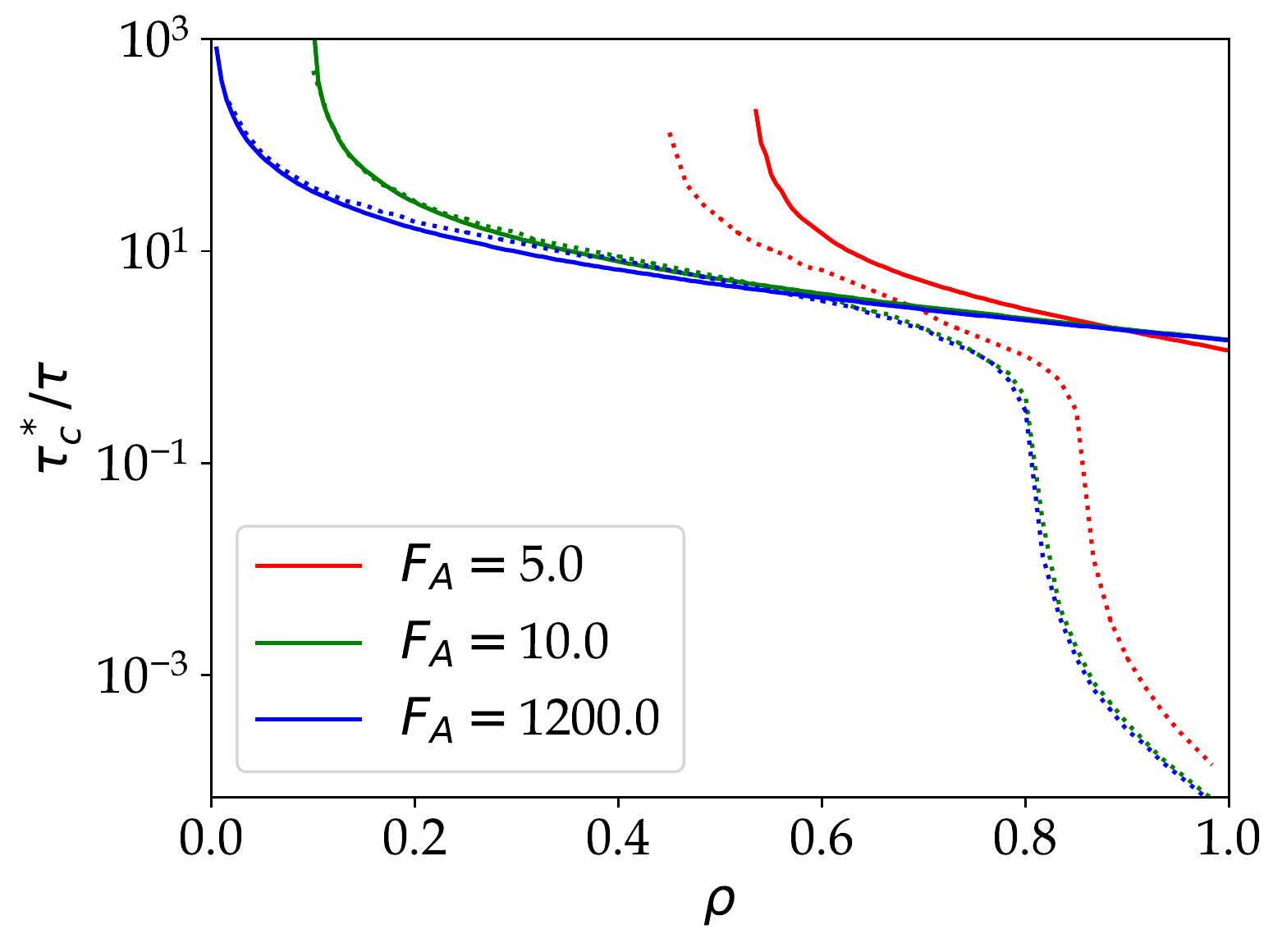}
\caption{Region of the non-monotony of the diffusion coefficient with $\tau_\alpha$ in the plane ($\rho$,$\tau^*/\tau$): above the curves, $D$ is a non-monotonic function of $\tau_\alpha$. At low density and strong active force, the scaling argument  {[Eq. \eqref{eq:D_simple}] (solid lines)} and the decoupling approximation {(dotted lines)} coincide. 
} 
\label{fig:NMD}
\end{center}
\end{figure}

\begin{figure}
\begin{center}
\includegraphics[width = 0.3\columnwidth]{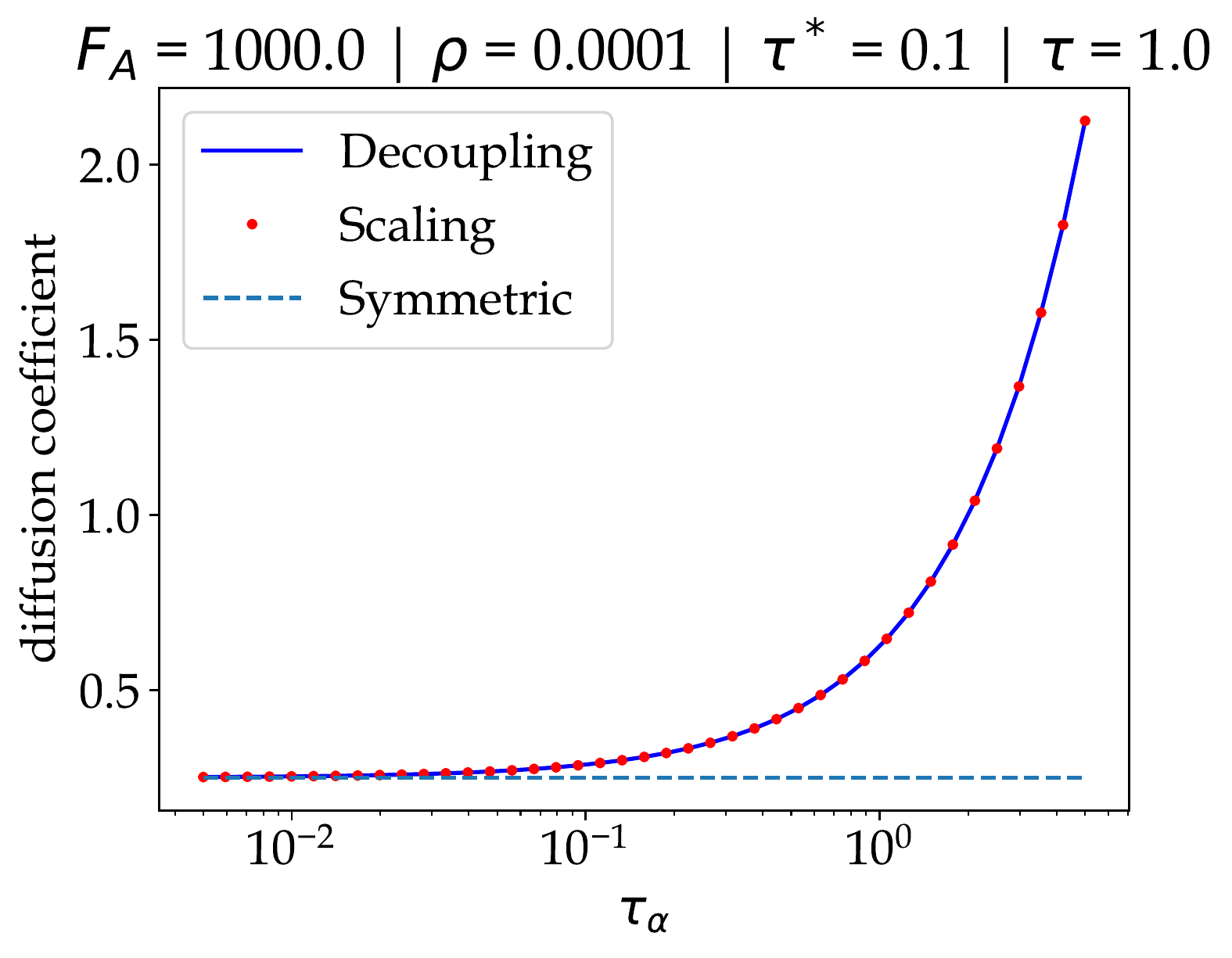}
\includegraphics[width = 0.3\columnwidth]{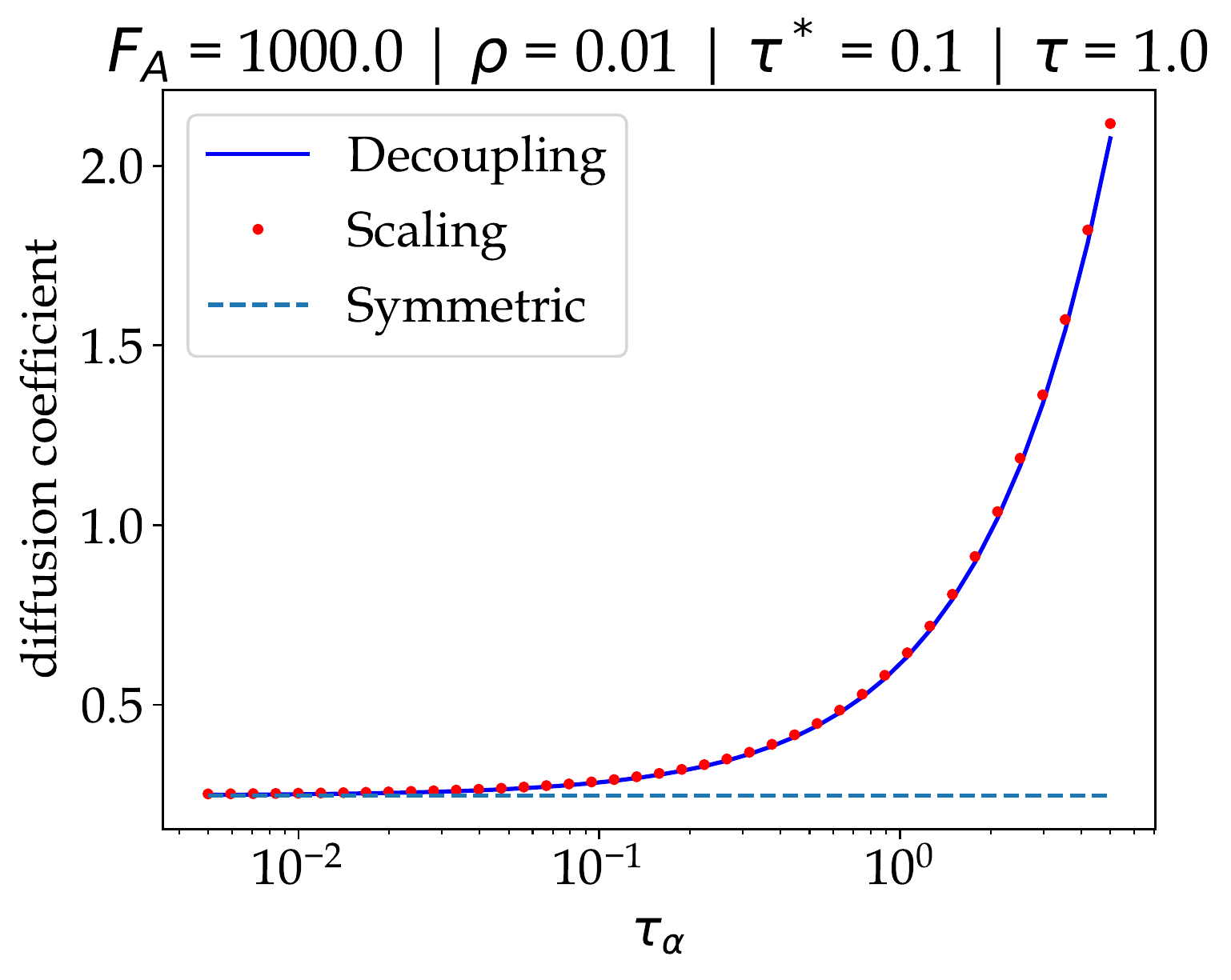}
\includegraphics[width = 0.3\columnwidth]{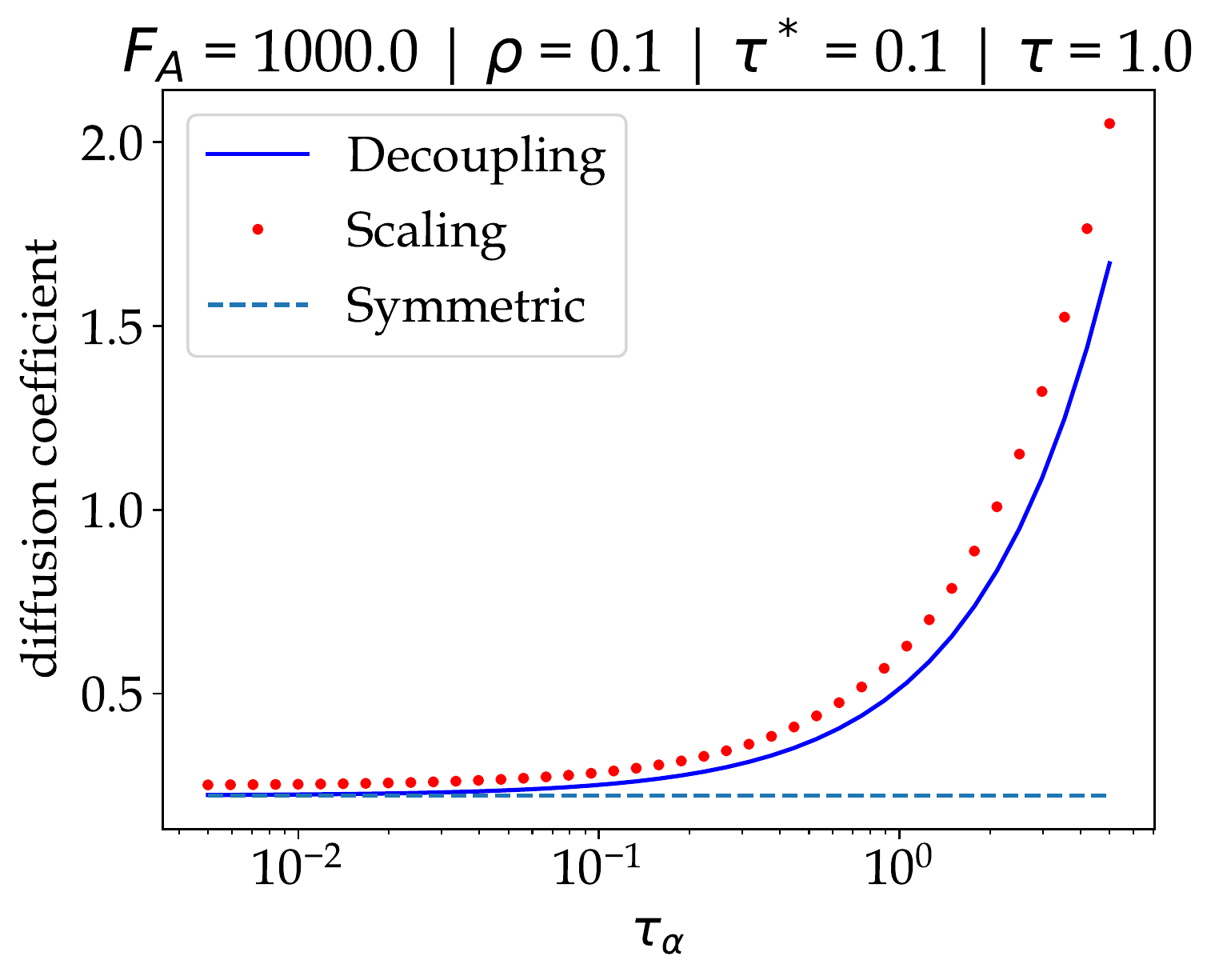}
\includegraphics[width = 0.3\columnwidth]{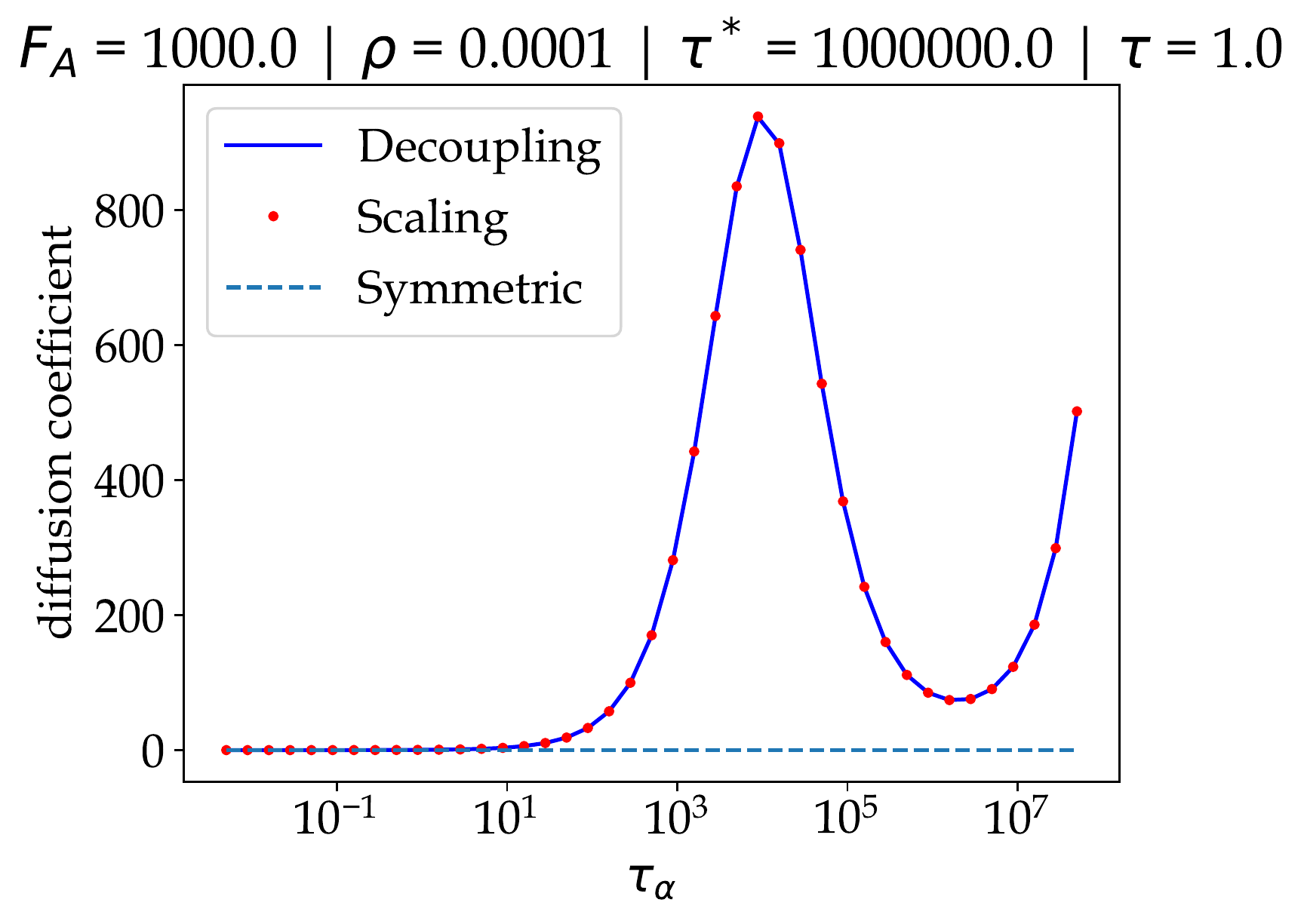}
\includegraphics[width = 0.3\columnwidth]{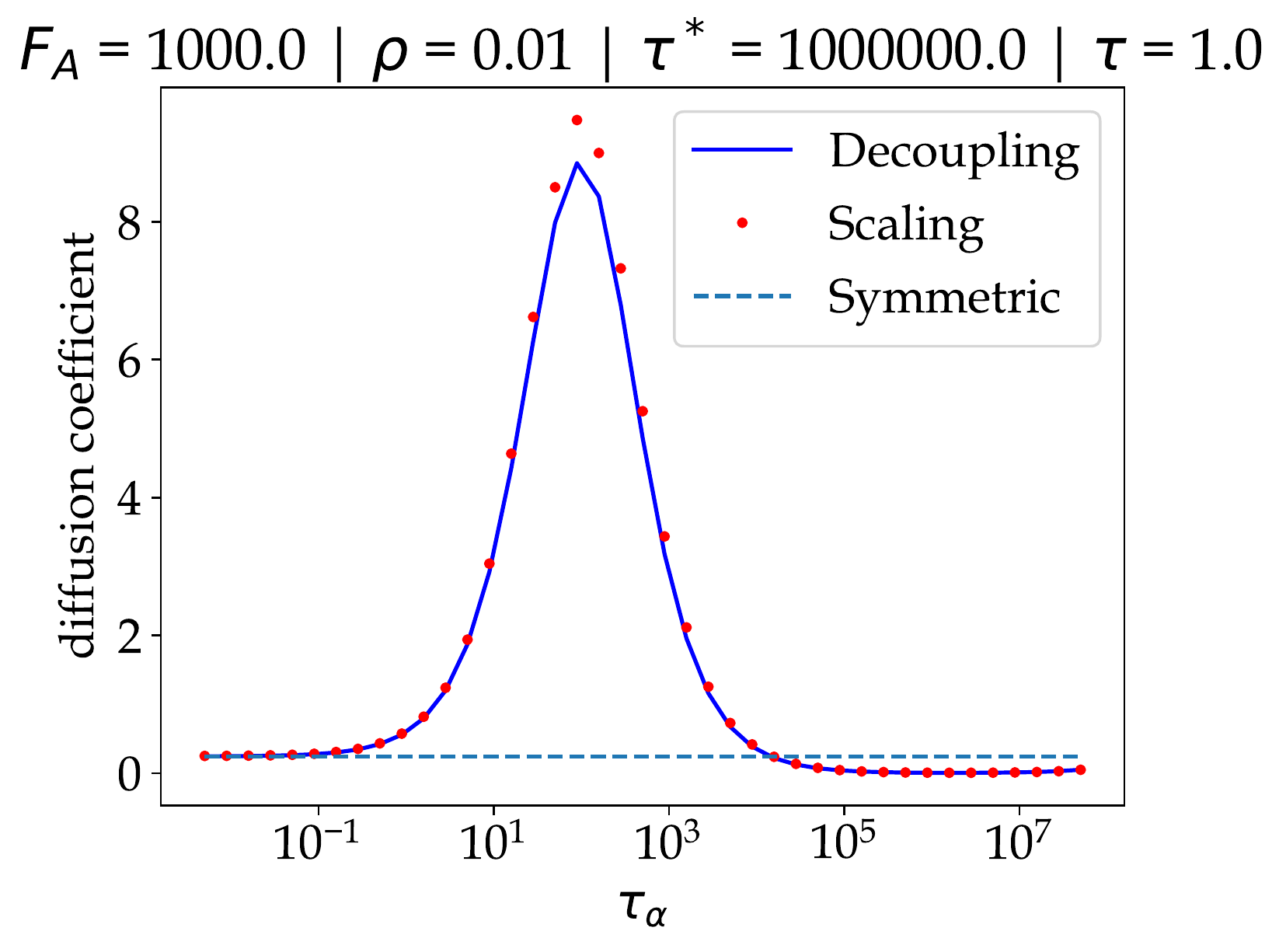}
\includegraphics[width = 0.3\columnwidth]{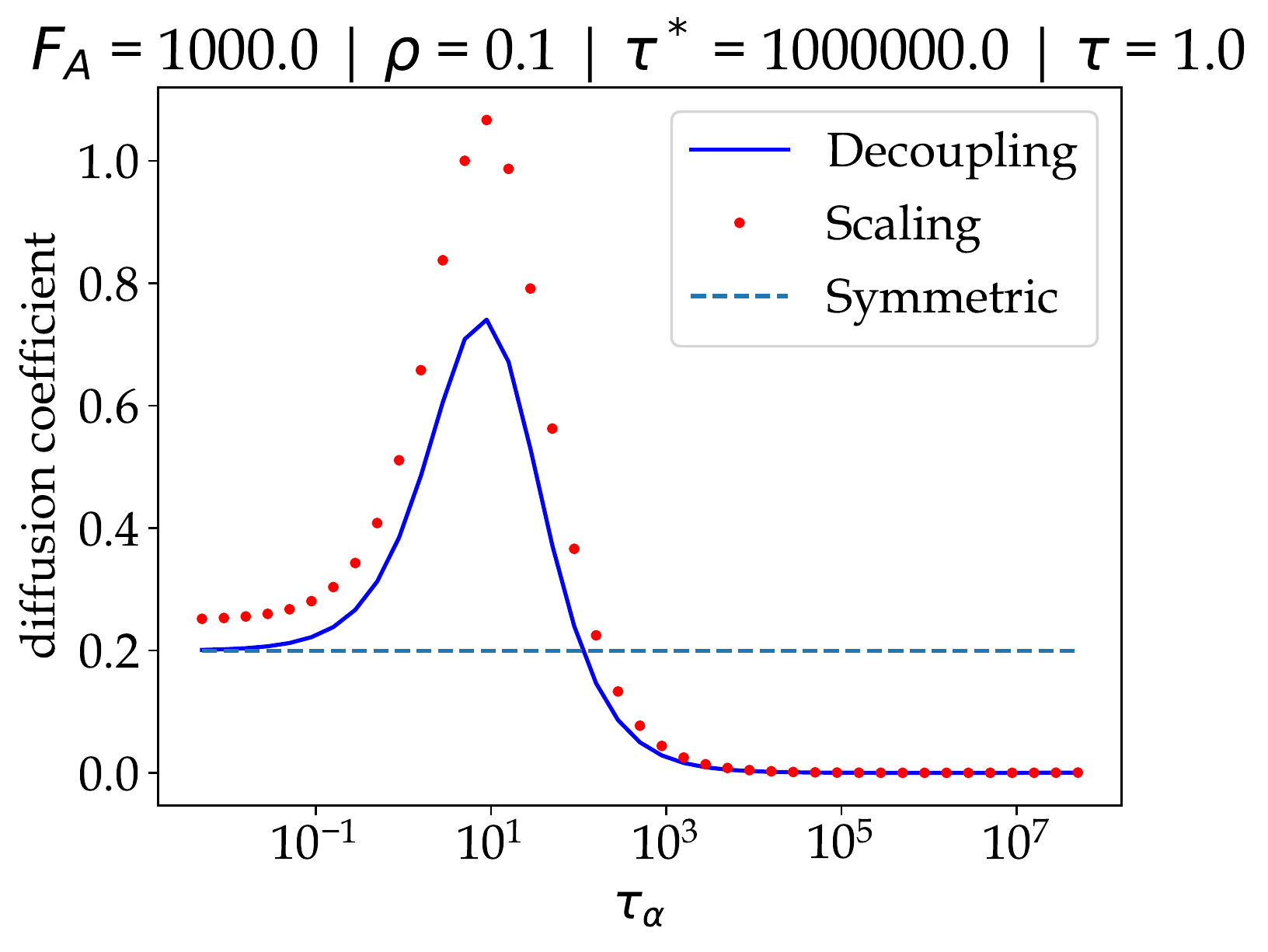}
\caption{Comparison of the diffusion coefficients computed by the two methods: decoupling approximation and scaling argument \blue{[Eq. \eqref{eq:D_simple}]}.} 
\label{fig:compare_decoup_simple}
\end{center}
\end{figure}

Studying this function of $\tau_\alpha$ in the case of an infinite force in dimension 2, we recover the scaling $\tau^* \geq \tau^*_c = 4\rho^{-1}$ for the condition of non monotony \cite{Bertrand2018a}. We determine the region of the non-monotony of the diffusion coefficient with $\tau_\alpha$ in the plane ($\rho$,$\tau^*/\tau$) (Fig. \ref{fig:NMD}). We also find that in the regime $\tau \ll \rho\tau^* \ll \tau^*$, the minimum of diffusion is reached for a tumbling rate close to the mobility of the obstacles (i.e. $\tau_\alpha \sim \tau^*$) and this minimum scales like $(\rho^{2}\tau^*)^{-1}$. In the case of a finite force $F_A \gg 1$, we find that if the density is below a critical value $\rho_c = \frac{16}{\tau(2+\ex{F_A/2})}$, then there is no more non monotony in $\tau_\alpha$, whatever the value of $\tau^*$.  This simple estimate therefore provides very accurate predictions of the diffusion coefficient in the low-density limit. The comparison between the estimate from Eq. \eqref{eq:D_simple} and the analytical approach presented in the main text is shown on Fig. \ref{fig:compare_decoup_simple}.

    \section{Low-density limit}

In the low-density limit,  the density profiles $h_\mu^{(\chi)}$ and the correlation functions $\gt_\mu^{(\chi)}$ are expanded as $h_\mu^{(\chi)} {=}_{\rho\to0} \rho h_{0,\mu}^{(\chi)} + \mathcal{O}(\rho^2)$ and $\gt_\mu^{(\chi)} {=}_{\rho\to0} \rho \gt_{0,\mu}^{(\chi)} + \mathcal{O}(\rho^2)$. The diffusion coefficient [Eq. (2) from the main text] is expanded as $D = D_0 + \rho \mathcal{D}_0 + \mathcal{O}(\rho^2)$, with
\begin{equation}
    D_0 = \frac{1}{2d\tau} +\frac{2d-1}{d}\frac{\tau^*}{\tau^2 \alpha}  [p_1^{(1)}-p_{-1}^{(1)}]^2, \label{eq:D0}\\
\end{equation}
\begin{equation}
    \mathcal{D}_0 = - \frac{1}{4d\tau} \sum_\chi \sum_{\epsilon=\pm1}p_\epsilon^{(\chi)}(1+h_{0,\epsilon}^{(\chi)}) + 2\epsilon p_\epsilon^{(\chi)} \gt_{0,\epsilon}^{(\chi)} -2\frac{2d-1}{d} \frac{\tau^*}{\tau^2\alpha} [p_1^{(1)}-p_{-1}^{(1)}] \sum_{\epsilon=\pm1} \epsilon p_\epsilon^{(1)}(1+h_{0,\epsilon}^{(1)}).\label{eq:DD0}
\end{equation}
In this limit, the density profiles $\boldsymbol{h}_0$ defined in the main text obey the linear equation: 
\begin{equation}
\label{eq:linearsyst_h0}
\boldsymbol{\mathcal{M}}_{\boldsymbol{h},0} \boldsymbol{h}_0 = \boldsymbol{x}_{\boldsymbol{h},0}
\end{equation}
with $\boldsymbol{\mathcal{M}}_{\boldsymbol{h},0} \equiv \boldsymbol{1} + \int_{\qq} \ex{-\ii q_1} \boldsymbol{M_0}^{-1}\boldsymbol{\Lambda_0} $ and $\boldsymbol{x}_{\boldsymbol{h},0} \equiv -\int_{\qq} \ex{-\ii q_1} \boldsymbol{M_0}^{-1} \boldsymbol{S_0}$
where $\boldsymbol{M_0}(\qq)$ is the $\rho\to0$ limit of $\boldsymbol{M}(\qq)$, with offdiagonal coefficients all equal to $\alpha/3$, and diagonal coefficients given by $[\boldsymbol{M_0}(\qq)]_{\chi\chi} = -\alpha + \sum_{\mu}   \left(\ex{-\ii\sigmaa\qq\cdot\ee_\mu} -1\right)a_\mu^{(\chi)}$, with $a_\mu^{(\chi)} = 1 + 4\frac{\tau^*}{\tau}p_\mu^{(\chi)}$. We recall the shorthand notation $\int_{\qq} = \int_{[-\pi,\pi]^d} \frac{\dd\qq}{(2\pi)^d}$ for the inverse Fourier transform. The expressions of the matrix $\boldsymbol{\Lambda_0}$ and of the vector $\boldsymbol{S_0}$  are
\begin{equation}
\boldsymbol{\Lambda_0} = \nonumber\\
\left(\begin{matrix} 
\left(\ex{\ii q_1} -1\right)a^{(1)}_1 & \left(\ex{-\ii q_1} -1\right)a^{(1)}_{-1} &
2\left(\cos q_2 -1\right)a^{(1)}_2 & 0 \\
\left(\ex{-\ii q_1} -1\right)a^{(1)}_1 & \left(\ex{\ii q_1} -1\right)a^{(1)}_{-1} &
2\left(\cos q_2 -1\right)a^{(1)}_2 & 0\\
\left(\ex{\ii q_2} -1\right)a^{(1)}_1 & \left(\ex{-\ii q_2} -1\right)a^{(1)}_{-1} &
2\left(\cos q_1 -1\right)a^{(1)}_2 & 0\\
\left(\ex{-\ii q_2} -1\right)a^{(1)}_1 & \left(\ex{\ii q_2} -1\right)a^{(1)}_{-1} &
2\left(\cos q_1 -1\right)a^{(1)}_2 & 0
\end{matrix}\right)
\end{equation}
and : 
\begin{equation}
\boldsymbol{S_0} =  2 \ii\dfrac{2 d \tau^*}{\tau} \left[p_1^{(1)} -p_{-1}^{(1)} \right]\left(\begin{matrix}
 \sin q_1\\
-\sin q_1\\
 \sin q_2\\
-\sin q_2
\end{matrix}\right)
\end{equation}


Similarly, it can be shown from Eq. \eqref{eqMG} that the vector  $\boldsymbol{\gt}_0 = (\gt_{0,1}^{(1)}, \gt_{0,-1}^{(1)},\gt_{0,2}^{(1)},\gt_{0,1}^{(2)})$ is the solution of the linear system $\boldsymbol{\mathcal{M}}_{\boldsymbol{\gt},0} \boldsymbol{\gt}_0 = \boldsymbol{x}_{\boldsymbol{\gt},0}$ with:
\begin{equation}
    \boldsymbol{\mathcal{M}}_{\boldsymbol{\gt},0}=\boldsymbol{1}+
\left(\begin{matrix}
\int_{\qq} \ex{-\ii\sigmaa q_1} \left[\boldsymbol{M_0}^{-1}\boldsymbol{\Lambda_{G_0}}\right]_1 \\
\int_{\qq} \ex{\ii\sigmaa q_1} \left[\boldsymbol{M_0}^{-1}\boldsymbol{\Lambda_{G_0}}\right]_1 \\
\int_{\qq} \ex{-\ii\sigmaa q_2} \left[\boldsymbol{M_0}^{-1}\boldsymbol{\Lambda_{G_0}}\right]_1 \\
\int_{\qq} \ex{-\ii\sigmaa q_1} \left[\boldsymbol{M_0}^{-1}\boldsymbol{\Lambda_{G_0}}\right]_3 
\end{matrix}\right) \\
\end{equation}
\begin{equation}
\boldsymbol{x}_{\boldsymbol{\gt},0} =-  \left(\begin{matrix}
\int_{\qq} \ex{-\ii\sigmaa q_1} \left[\boldsymbol{M_0}^{-1}\left(\dfrac{1}{\rho}\boldsymbol{X}\cdot \boldsymbol{H} + \boldsymbol{E_0} + \boldsymbol{F_0}\right)\right]_1 \\
\int_{\qq} \ex{\ii\sigmaa q_1} \left[\boldsymbol{M_0}^{-1}\left(\dfrac{1}{\rho}\boldsymbol{X}\cdot \boldsymbol{H} + \boldsymbol{E_0} + \boldsymbol{F_0}\right)\right]_1 \\
\int_{\qq} \ex{-\ii\sigmaa q_2} \left[\boldsymbol{M_0}^{-1}\left(\dfrac{1}{\rho}\boldsymbol{X}\cdot \boldsymbol{H} + \boldsymbol{E_0} + \boldsymbol{F_0}\right)\right]_1 \\
\int_{\qq} \ex{-\ii\sigmaa q_1} \left[\boldsymbol{M_0}^{-1}\left(\dfrac{1}{\rho}\boldsymbol{X}\cdot \boldsymbol{H} + \boldsymbol{E_0} + \boldsymbol{F_0}\right)\right]_3 
\end{matrix}\right)
\end{equation}
where $[ \cdot ]_i$ means that we take the $i$-th line, and with
%
\begin{equation}
\boldsymbol{M_0} =  \left(\begin{matrix}
-\alpha+\sum_\mu(\ex{-\ii q_\mu}-1)a_\mu^{(1)} & \alpha/3 & \alpha/3 &\alpha/3  \\
 \alpha/3 & -\alpha+\sum_\mu(\ex{-\ii q_\mu}-1)a_\mu^{(-1)} & \alpha/3 &\alpha/3  \\
 \alpha/3 & \alpha/3 &-\alpha+\sum_\mu(\ex{-\ii q_\mu}-1)a_\mu^{(2)} & \alpha/3  \\
\alpha/3 & \alpha/3 & \alpha/3 &-\alpha+\sum_\mu(\ex{-\ii q_\mu}-1)a_\mu^{(-2)} & 
\end{matrix}\right), 
\end{equation}
\begin{equation}
\boldsymbol{E_0} = \dfrac{1}{\rho}\sigmaa\left(\begin{matrix}
H^{(1)}\left((a_1^{(1)}-1) (\ex{-\ii \sigmaa	q_1}-1) - (a_{-1}^{(1)}-1) (\ex{\ii \sigmaa	q_1}-1)\right)\\
H^{(-1)}\left((a_{-1}^{(1)}-1) (\ex{-\ii \sigmaa	q_1}-1) - (a_1^{(1)}-1)  (\ex{\ii \sigmaa	q_1}-1)\right)\\
-2 \ii (a_2^{(1)}-1) H^{(2)} \sin(\sigmaa	q_1) \\
-2 \ii (a_2^{(1)}-1) H^{(-2)} \sin(\sigmaa	q_1) \\
\end{matrix}\right),
\end{equation}
\begin{align}
\boldsymbol{\Lambda_{G_0}} = & \left(\begin{matrix}
\left(\ex{\ii\sigmaa q_1} -1\right)a_1^{(1)} & \left(\ex{-\ii\sigmaa q_1} -1\right)a_{-1}^{(1)} & 2\left(\cos\left(\sigmaa q_2\right) -1\right)a_2^{(1)} & 0 \\
-\left(\ex{-\ii\sigmaa q_1} -1\right)a_1^{(1)} & -\left(\ex{\ii\sigmaa q_1} -1\right)a_{-1}^{(1)} & -2\left(\cos\left(\sigmaa q_2\right) -1\right)a_2^{(1)} & 0 \\
0 & 0 & 0 & 2\ii a_2^{(1)} \sin(\sigmaa q_1) \\
0 & 0 & 0 & 2\ii a_2^{(1)} \sin(\sigmaa q_1)
\end{matrix}\right), 
\end{align}
and
\begin{equation}
\boldsymbol{F_0} = -\sigmaa\left(\begin{matrix}
(a_1^{(1)}-1)\left(\ex{-\ii\sigmaa q_1} + \dfrac{1}{\rho}h_1^{(1)}\right) - (a_{-1}^{(1)}-1)\left(\ex{\ii\sigmaa q_1} + \dfrac{1}{\rho}h_{-1}^{(1)}\right) \\
(a_{-1}^{(1)}-1)\left(\ex{-\ii\sigmaa q_1} + \dfrac{1}{\rho}h_1^{(-1)}\right) - (a_{1}^{(1)}-1)\left(\ex{\ii\sigmaa q_1} + \dfrac{1}{\rho}h_{-1}^{(-1)}\right) \\
(a_2^{(1)}-1)\left(\ex{-\ii\sigmaa q_1} + \dfrac{1}{\rho}h_1^{(2)}\right) - (a_{2}^{(1)}-1)\left(\ex{\ii\sigmaa q_1} + \dfrac{1}{\rho}h_{-1}^{(2)}\right) \\
(a_2^{(1)}-1)\left(\ex{-\ii\sigmaa q_1} + \dfrac{1}{\rho}h_1^{(2)}\right) - (a_{2}^{(1)}-1)\left(\ex{\ii\sigmaa q_1} + \dfrac{1}{\rho}h_{-1}^{(2)}\right).
\end{matrix}\right)
\end{equation}

\section{Comparison with the discrete-time approach in the low-density limit}

\begin{figure}
    \centering
    \includegraphics[width=8cm]{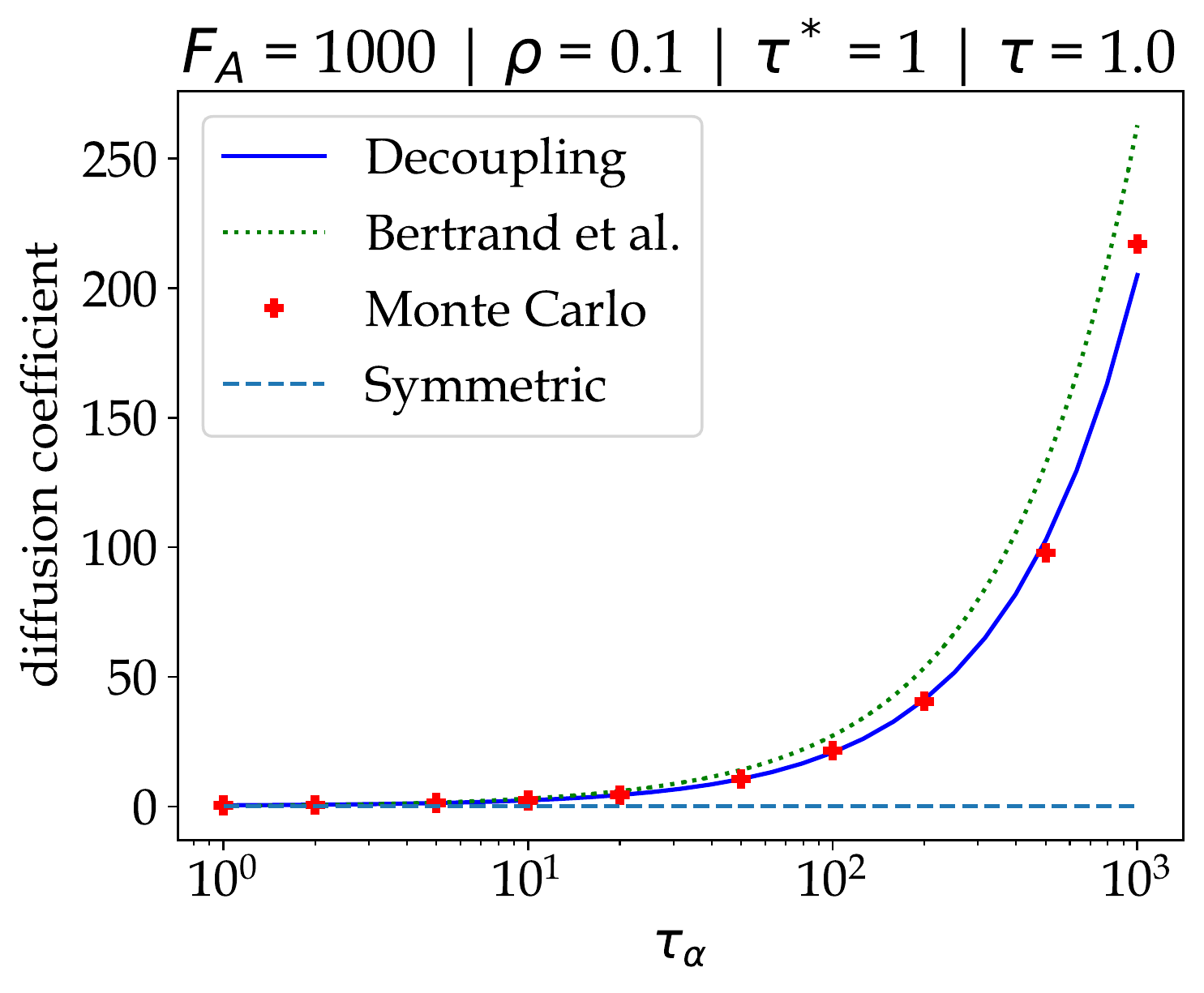}
    \includegraphics[width=8cm]{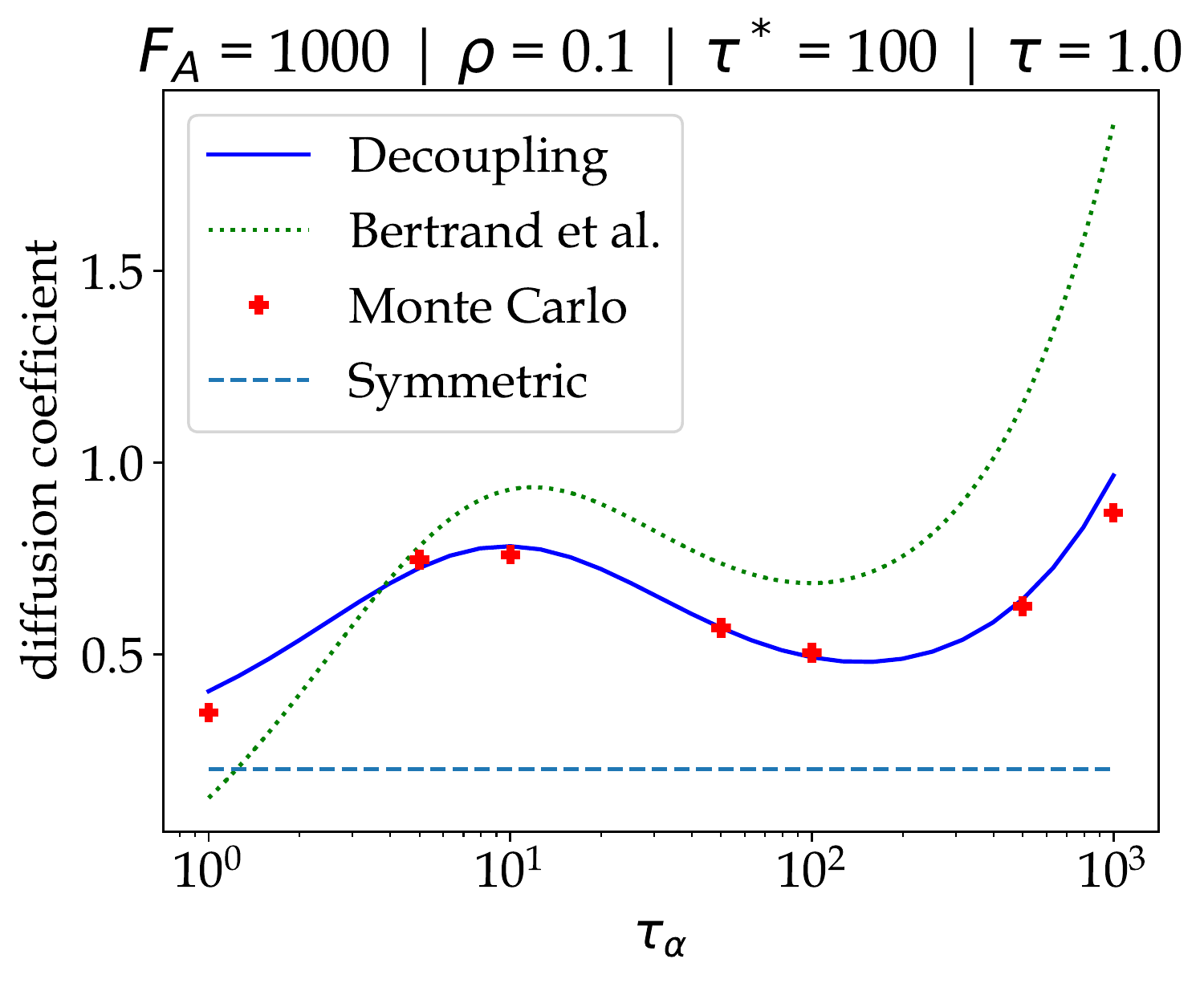}
    \caption{Comparison between the result from our decoupling approximation and the discrete-time approach presented in Ref. \cite{Bertrand2018}.}
    \label{fig:compare_Bertrand}
\end{figure}

We compare the results from our decoupling approximation with the results from Ref. \cite{Bertrand2018a} on Fig. \ref{fig:compare_Bertrand} (see main text for discussion).

\section{High-density limit} 

In the high-density limit $\rho\to1$, writing the  expansions  $h_\mu^{(\chi)} {=}_{\rho\to1} (1- \rho) h_{1,\mu}^{(\chi)} + \mathcal{O}[(1-\rho)^2]$, $\gt_\mu^{(\chi)} {=}_{\rho\to1} (1-\rho) \gt_{1,\mu}^{(\chi)} + \mathcal{O}[(1-\rho)^2]$, and using Eq. (2) yields the asymptotic form of the diffusion coefficient $D {=}_{\rho\to 1} (1-\rho) \mathcal{D}_1 + \mathcal{O}[(1-\rho)^2]$ with
\begin{equation}
\mathcal{D}_1 = \frac{1}{4d\tau} \sum_\chi \sum_{\epsilon=\pm1}p_\epsilon^{(\chi)}(1-h_{1,\epsilon}^{(\chi)}) - 2\epsilon p_\epsilon^{(\chi)} \gt_{1,\epsilon}^{(\chi)}.
\end{equation}
The vectors $\boldsymbol{h}_1 = (h_{1,1}^{(1)}, h_{1,1}^{(-1)},h_{1,1}^{(2)})$ and $\boldsymbol{\gt}_1 = (\gt_{1,1}^{(1)}, \gt_{1,1}^{(-1)},\gt_{1,1}^{(2)},\gt_{1,2}^{(1)})$, can be calculated after a linearization of Eqs. (5) and (6) in the limit $\rho\to1$.

Interestingly, in the high-density regime, the matrix $\boldsymbol{M}(\qq)$ involved in Eqs. (5) and (6) is even simpler than in the low-density regime, and has all off-diagonal coefficients equal to $-\alpha/(2d-1)$, and all diagonal coefficients equal to $-\alpha +2d[\lambda(\qq)-1]$ where  $\lambda(\qq) = \sum_\mu \frac{1}{2d}\ex{-\ii q_\mu}$ is the structure factor associated to the random walk of an isolated and symmetric random walker \cite{Hughes1995}.The high density limit of the matrix $\boldsymbol{M}(\qq)$ can be reduced by noticing that its eigenvalues are $2d ( \lambda(\qq) - 1)$, associated to the eigenvector $(1)_{i\in \llb 1,2d\rrb}$ and $2d [ \lambda(\qq) - 1 - \alpha/(2d-1)]$ associated to the space orthogonal to $(1)_{i\in \llb 1,2d\rrb}$.

The vectors $\boldsymbol{h}_1$ and $\boldsymbol{\gt}_1$ are then written as the solutions  of linear systems $\boldsymbol{\mathcal{M}}_{\boldsymbol{h},1} \boldsymbol{h}_1 = \boldsymbol{x}_{\boldsymbol{h},1}$ and $\boldsymbol{\mathcal{M}}_{\boldsymbol{\gt},1} \boldsymbol{\gt}_1 = \boldsymbol{x}_{\boldsymbol{\gt},1}$, with
\begin{equation}
\boldsymbol{\mathcal{M}}_{\boldsymbol{h},1} = 
\left(\begin{matrix}
2d - c + bp_1\frac{2d\tau^*}{\tau} & b - c - bp_{-1}\frac{2d\tau^*}{\tau} & 0 \\
b - c - bp_1\frac{2d\tau^*}{\tau} & 2d -c + bp_{-1}\frac{2d\tau^*}{\tau} & 0 \\
1 & 1 & 2d-2
\end{matrix}\right)
\end{equation}
 \begin{equation}
\boldsymbol{x}_{\boldsymbol{h},1} = 
\dfrac{2d\tau^*}{\tau} (p_1^{(1)}-p_{-1}^{(1)}) \left(\begin{matrix}
 b\\
 -b\\
 0
\end{matrix}\right)
\end{equation}
\begin{align}
\boldsymbol{\mathcal{M}}_{\boldsymbol{\gt},1} \equiv & 
\left(\begin{matrix}
2d & 2d & 4d^2-4d & 0 \\
2d-1 & -1 & -(2d-2) & 0 \\
-1 & 2d-1  & -(2d-2) & 0 \\
0 & 0 & 0 & 1 \\
\end{matrix}\right) -\left(\begin{matrix}
b_0 &b_0 & (2d-2) b_0 & 0 \\
\dfrac{2da-b}{2d} & \dfrac{(2d-1)b - 2da}{2d} & -b\dfrac{2d-2}{2d} & 0 \\
\dfrac{(2d-1)b - 2da}{2d} & \dfrac{2da-b}{2d} & -b\dfrac{2d-2}{2d} & 0 \\
a - c & c - a & 0 & (2d-2) a - b -(2d-4) c \\
\end{matrix}\right) \nonumber\\
&+ \dfrac{2d \tau^*}{\tau} \left(	\begin{matrix}
2p_1b_0 & 2 p_{-1} b_0 & (4d-4)p_2b_0 & 0 \\
\dfrac{2d-2}{2d} p_1 b & \dfrac{2d-2}{2d} p_{-1} b & -\dfrac{4 d - 4}{2d} p_2 b & 0 \\
\dfrac{2d-2}{2d} p_1 b & \dfrac{2d-2}{2d} p_{-1} b & -\dfrac{4 d - 4}{2d} p_2 b & 0 \\
0 & 0 & 0 & 0 \\
\end{matrix}\right)	 
\end{align}
\begin{equation}
\boldsymbol{x}_{\boldsymbol{\gt},1} = -\dfrac{2d\tau^*}{\tau}\left( \begin{matrix}
0 \\ 
p_1(1-h_1^{(1)})(\beta - b) - p_{-1}(1+h_1^{(1)})\beta \\
p_{-1}(1+2_1^{(1)})(\beta - b) - p_{1}(1-h_1^{(1)})\beta \\
\left(p_1(1-h_1^{(1)}) - p_{-1}(1+h_1^{(1)})\right)(\beta-c) \\
\end{matrix}\right)
+ \dfrac{2d\tau^*}{\tau}(1-(p_1 - p_{-1})h_1^{(1)})\left(\begin{matrix}
b_0 \\ 
-\dfrac{b}{2d} \\
-\dfrac{b}{2d} \\
0 \\
\end{matrix}\right)
\end{equation}
The expressions involve the following quantities:
\begin{align}
a & = \zeta [\widehat{Q}(\zz|\zz;\zeta)-\widehat{Q}(\ee_1|\zz;\zeta)]\\
b & = \zeta [\widehat{Q}(\zz|\zz;\zeta)-\widehat{Q}(2\ee_1|\zz;\zeta)]\\ 
b_0 & = \widehat{Q}(\zz|\zz;1)-\widehat{Q}(2\ee_1|\zz;1)\\
c & = \zeta\left[\widehat{Q}\left(\zz \left\lvert\zz ; \zeta\right.\right) - \widehat{Q}\left(\ee_1 + \ee_\mu \left\lvert\zz ; \zeta\right.\right)\right] \\ 
\beta & = \zeta\widehat{Q}\left(\zz \left\lvert\zz ; \zeta\right.\right) 
\end{align}
 where $\zeta = \frac{2d-1}{2d-1+\alpha}$. Remarkably, they only depend on $\widehat{Q}(\rr|\zz;\zeta) = \int_{\qq} \ex{-\ii \qq\cdot\rr}/[1-\zeta\lambda(\qq)]$, which  is the
generating function associated with the propagator of an isolated random walk starting from $\zz$ and arriving at site $\rr$ on a $d$-dimensional lattice \cite{Hughes1995}.


Note that the solution of the system $\boldsymbol{\mathcal{M}}_{\boldsymbol{h},1} \boldsymbol{h}_1 = \boldsymbol{x}_{\boldsymbol{h},1}$ is remarkably simple:
\begin{align}
h^{(1)}_{\pm 1} &\underset{\rho\to 1}{=} \pm  (1-\rho) \dfrac{\dfrac{2d\tau^*}{\tau} b \left(p^{(1)}_{1}-p^{(1)}_{-1}\right)}{\dfrac{2d\tau^*}{\tau} b \left(p^{(1)}_{1}+p^{(1)}_{-1}\right) + 2d - b} + \mathcal{O}[(1-\rho)^2], \\
h^{(1)}_{\pm 2} &\underset{\rho\to 1}{=} \mathcal{O}[(1-\rho)^2].
\end{align}

\section{Numerical resolution}

As we are interested only in the final value of the diffusion coefficient, we use a fixed point method inspired by the Euler method in order to solve equations (3) and (4) from the main text in the stationary regime. These equations are of the form :
\begin{equation}
0 = f(h,g).
\end{equation} 
To find the root of this function $f$, we compute numerically the sequence defined by the following equation, where $s$ is a linear invertible application : 
\begin{align}
(h_{n+1},g_{n+1}) = (h_n, g_n) + s \cdot f(h_n,g_n)
\end{align}
with initial values $(h_0)^{(\chi)}_{\rr} = \rho$ and $(g_0)^{(\chi)}_{\rr} = 0$.
Vectors $h_n$ and $g_n$ are infinite dimensional and therefore cannot be represented numerically. To deal with this issue we use the fact that $h^{(\chi)}_{\rr}$ and $g^{(\chi)}_{\rr}$ vanish when $\rr$ goes to infinity. It allows us to impose a cut-off $L \in \mathbb{N}$ such that we consider $h^{(\chi)}_{\rr} = g^{(\chi)}_{\rr} = 0$ when $\rr$ has a component greater than $L$. We verify that for $L$ large enough, the values obtained do not depend on $L$. Note that the complexity in $L$ is $O(L^2)$ (resolution in two dimensions).

For the choice of $s$, the exponential rate of convergence is equal to the highest eigenvalue of $c = s \cdot \dd f (h,g) - \textbf{1}$ where $(h,g)$ here denote the root of $f$. If we choose $s = k \textbf{1}$, we get the Euler method with time step $k$. We see that if $k$ is too small, eigenvalues of $c$ will be too close to 1 and the convergence will be too slow, and if it is too large, we will have eigenvalues greater than 1 and the sequence will diverge. We use the following formula for $k$, which enables a quick convergence and ensures the sequence will not diverge : 
\begin{equation}
k = \dfrac{1}{\dfrac{1}{\tau_\alpha} + \dfrac{1}{\tau} + \dfrac{1}{\tau^*}}
\end{equation}
We can control the error of the method by evaluating the difference between two consecutive terms of the sequence.

\section{Monte-Carlo simulations}

We consider a $d-$dimensional lattice with $M$ sites and initialize the $N$ bath particles in a random configuration, with density $\rho = N/M$. In the case $d = 2$, we used a square lattice with $M = 200 \times 200$ sites, with periodic boundary conditions in both directions, and we checked that results are independent of the box size. For $d = 3$, we used a cubic lattice with $M=30 \times 30 \times 30$ sites, with periodic boundary conditions. In the case of a two-dimensional strip-like lattice, we used $M=2000 \times 3$ sites. Numerical results reported in the main text are obtained from averages over about $10^3 - 10^4$ realizations.

\bibliographystyle{apsrev4-1}
\bibliography{/Users/pierreillien/work/docs/library}